\documentclass[a4paper,12pt]{article}
\usepackage[top=3cm, bottom=3cm, left=3cm, right=3cm]{geometry} 
\usepackage{color}
\usepackage{rotating}
\usepackage{graphicx}
\usepackage{amsmath}
\usepackage{amssymb}
\usepackage{float}
\usepackage{subfig}
\usepackage{subfloat}
\usepackage{hyperref}
\def\displayandname#1{\rlap{$\displaystyle\csname #1\endcsname$}%
                      \qquad \texttt{\char92 #1}}

\begin{document}

\title{\bf{Primary Ionization and Particle Identification with Straw Tube Detectors}}

\author{R. Kanishka\thanks{email:
kanishka.rawat.phy@gmail.com}~ \\
\it $^*$Department of Physics, University Institute of Sciences, \\Chandigarh University, Mohali, Punjab, India 140413. \\
}
\maketitle

\begin{abstract}{The charged particles are tracked in the high energy physics detectors to provide the information of their properties. One of the tracking detector is straw tube detector that has been used by many experiments. The motivation behind the current work is to study the primary ionization and particle identification using straw tube detectors. Additionally, we report the decay of $^{60}$CO for the study of gamma peaks since these are used in cobalt therapy, that is beneficial for cancer/tumor treatment. The various studies like primary ionization, spatial co-ordinate distributions in the different gas mixtures, transition radiation, drift velocities of electrons and diffusion coefficients using different xenon-based gas mixtures have been obtained. These studies have been done after the optimization of xenon-based gas mixtures for a deeper understanding. The gas mixture that shows maximum transition radiation among the xenon-based gas mixtures was found to be $Xe:He:CH_{4}$ :: 30:55:15. The gas mixture that posses maximum primary ionization has been observed to be $Xe:CO_{2}$ :: 70:30. Finally, the simulations have been carried out for the particle identification in the straw tube detectors with different particles i.e., muons, pions and kaons.}
\end{abstract}

\newpage

\section{Introduction}
\label{sec:intro}

The gaseous detectors have been used in many particle physics experiments since a long time. In order to study the particle properties in the gaseous detectors the optimization of various parameters like detector geometries and/or gas mixtures are crucial. The straw tube detectors are used by many ongoing and upcoming particle physics experiments to track the different particles. The COMPASS \cite{compass}, PANDA \cite{pandas1}--\cite{pandas3}, ATLAS \cite{atlas0}--\cite{atlas5}, NA62\cite{na62}, GlueX\cite{gluex} and an upcoming experiment like DUNE \cite{dune1}--\cite{dune2} have been using the straw tube trackers (STT) \cite{straw1}. Each of these experiments have different physics goals and hence posses the different geometries and, composition of gas mixtures. The xenon-based gas mixture is crucial in transition radiation tracker (TRT) as these have tracking capabilities along with particle identification and also absorb the transition radiation photon. The straw tube detectors are important for TRT as these have a high degree of modularity of the detector and can be easily integrated into a medium producing transition radiation and absorb them. There are numerous characteristics of straw tube detectors such as excellent vertex, momentum, angular, and time resolution that makes them special for doing several studies.

The current work aims to do the primary ionization and particle identification along with the various studies using straw tube detectors. A charged particle when passes the gaseous medium ionizes the gas atoms called primary ionization. The primary ionization once reaches the Raether limit then the avalanche progresses towards the discharge. These discharges create a spark and damages the detector, making them non-operational \cite{rout}. Hence repairing these detectors again would be time consuming and costly. In order to prevent the detectors from the damage a detailed study of primary ionization is crucial. The xenon-based gas mixtures have been chosen for the study as xenon shows highest ionization compared to the other noble gases. The quenchers have also been added in the presented studies. The gas mixtures have been optimized to select the best gas mixture for primary ionization. The two new gases also have been introduced, namely $Xe:CO_{2}:N_{2}$ and $Xe:C_{5}H_{12}$. Additionally, the gas properties such as drift velocities, longitudinal and transverse diffusion coefficients have also been obtained with different xenon-based gas mixtures. In addition to this work, the straw tube detector was chosen to study x-ray transition radiation (XTR) \cite{straw2} which is caused when a charged particle passes through the boundary between two materials with different dielectric constants. The multiple foils interleaved with air have been simulated in the detector for detecting the XTRs. In this paper, the particle identification has also been done using ionization process by simulating the different particles namely, muons, pions and kaons in the xenon-based gas mixture. The straw tube detector have also been used to study the decay of $^{60}$CO, that gives gamma peaks which is useful for cobalt therapy. The cobalt therapy is helpful in the cancer/tumor treatment. It is a novel concept in the STT detector.

The paper has been organized as follows: section~\ref{sec:num} describes different studies with various gas mixtures. The results have been discussed in the section~\ref{sec:results}. The summary and conclusions have been discussed in the section~\ref{sec:summ}.

\section{Studies with Various Gas Mixtures}
\label{sec:num}

\begin{figure}[htbp]
\centering
\includegraphics[width=1.\textwidth]{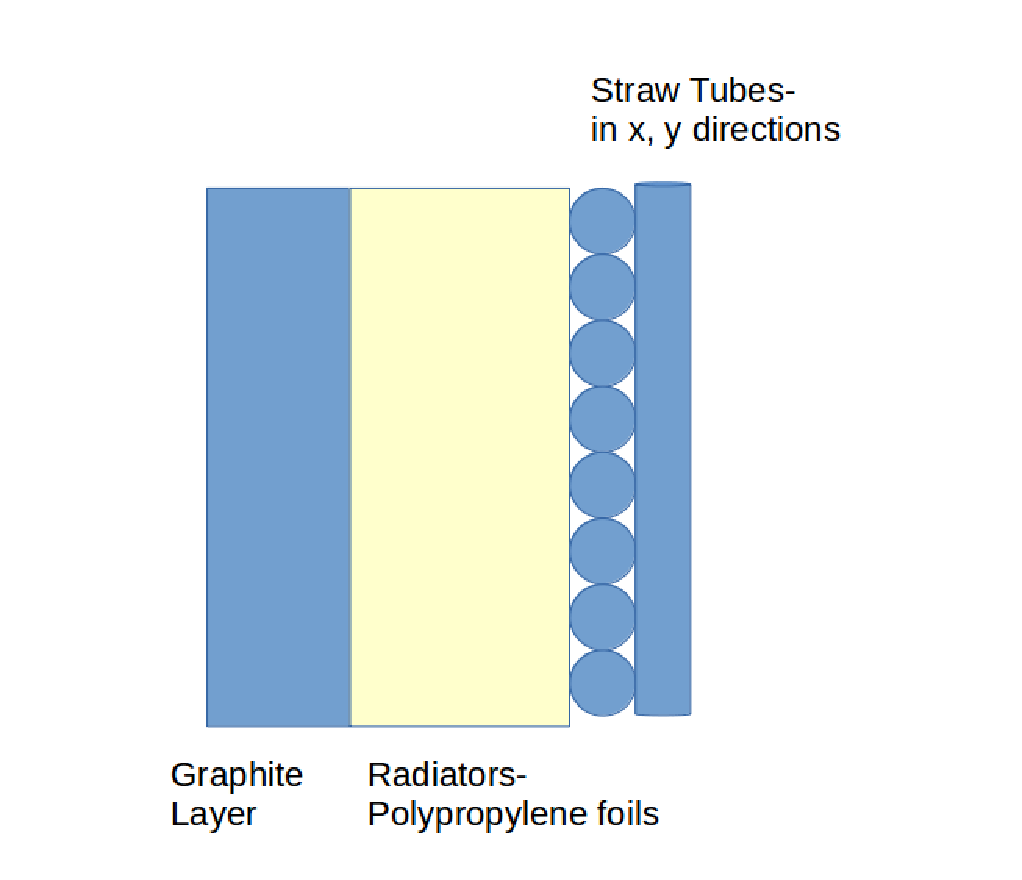}

\caption{The schematic of the simulated detectors.\label{fig:O}}
\end{figure}

The straw tube detector has been simulated using geant4 toolkit \cite{geant4} and the data was further analyzed using ROOT software \cite{root}. The set up consists of firstly a graphite layer and then the radiators that are made up of polypropylene foils. These foils have 18 $\mu$m thickness and 117 $\mu$m air gap between them \cite{dune2}. The straw tubes are placed across the x-y plane. The straw tubes are made up of mylar and the inner wire is made up of 95\% tungsten and 5\% gold. These straw tubes are filled with xenon-based gas mixtures as mentioned in the table~\ref{table1}. Table~\ref{table2} shows the dimensions of the detector that have been used for the analysis. The figure~\ref{fig:O} shows the schematic of simulated detector.

\begin{table}[htbp]
\centering
\caption{The different gas mixtures and their compositions (Comp.) used for the study \cite{straw3}--\cite{straw8}.\label{table1}}
%\smallskip
\begin{tabular}{|c|c|c|c|c|c|c|c|}
\hline
Gases & Comp. 1 & Comp. 2 & Comp. 3 & Comp. 4 & Comp. 5 & Comp. 6 & Comp. 7 \\
\hline
$XeCO_{2}O_{2}$ & 80:10:10 & 70:20:10 & 70:27:3 & 60:20:20 & 50:30:20 & 40:50:10 & 30:55:15\\
$XeCO_{2}N_{2}$ & 80:10:10 & 70:20:10 & 70:27:3 & 60:20:20 & 50:30:20 & 40:50:10 & 30:55:15\\
$XeHeCH_{4}$ & 80:10:10 & 70:20:10 & 70:27:3 & 60:20:20 & 50:30:20 & 40:50:10 & 30:55:15\\
$XeCO_{2}$ & 90:10 & 80:20 & 70:30 & 60:40 & 50:50 & 40:60 & 30:70 \\
$XeC_{5}H_{12}$ & 90:10 & 80:20 & 70:30 & 60:40 & 50:50 & 40:60 & 30:70 \\
$XeCH_{4}$ & 90:10 & 80:20 & 70:30 & 60:40 & 50:50 & 40:60 & 30:70 \\
\hline
\end{tabular}
\end{table}

\begin{table}[htbp]
\centering
\caption{The dimensions of the straw tube detector.\label{table2}}
%\smallskip
\begin{tabular}{|c|c|}
  \hline
Dimensions of graphite target & 1m $\times$ 1m $\times$ 0.004m\\
Dimensions of polypropylene foils & 1m $\times$ 1m $\times$ 3cm\\ 
No. of Straw tubes & 100\\
Straw tube inner radius & 2.5 mm\\
Straw tube outer radius & 3.5 mm\\
Wire inner radius & 5 $\mu$ m\\
Wire outer radius & 10 $\mu$ m\\
Length of wire & 900 mm \\
\hline
\end{tabular}
\end{table}

The gas mixtures have been chosen by selecting the base component xenon that posses excellent x-ray absorption as well as primary ionization \cite{straw12b}. The xenon gas has been combined with the different quenchers as mentioned in the table~\ref{table1}. The optimization of these gas mixtures with different ratios have been done for the study of primary ionization, XTR and gas properties that has been discussed in the next section. The gas mixtures have been taken at STP for the analysis. Now we discuss the results in the next section.

\section{Results}
\label{sec:results}

For the various studies, 10,000 different particles i.e., $^{60}$CO source, pions, kaons, muons, electrons were shot in the z-direction in the simulated detector using geant4. It was further analyzed that has been discussed in the next sub-sections.

\subsection{$^{60}$CO Decay}

\begin{figure}[htbp]
\centering
\includegraphics[width=1.\textwidth]{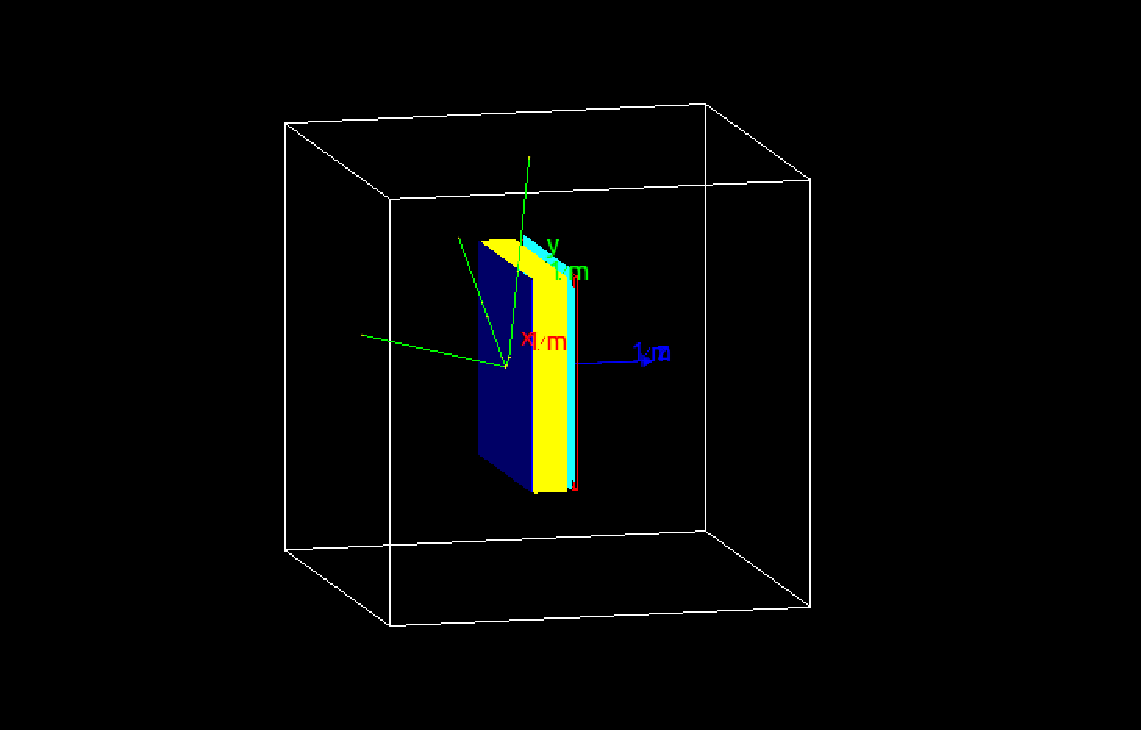}

\caption{The geant4 display for straw tube detectors irradiated with the $^{60}$CO source. Note: The blue plate represents graphite plate, yellow foils represents transition radiation detectors, cyan tubes represents straw tubes along x- and y- directions both, red box represents the gas volume, white box represents the world volume. The green colour particles are the gammas and yellow points represents the hits due to interaction of particles with the gas mixture.\label{fig:One}}
\end{figure}

\begin{figure}[htbp]
\centering
\includegraphics[width=1.\textwidth]{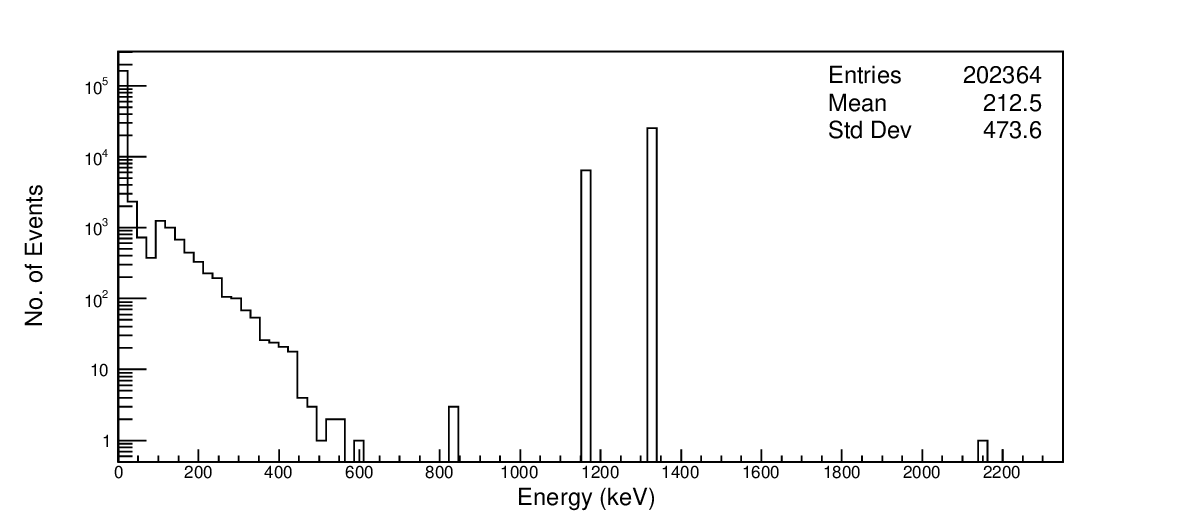}
\includegraphics[width=1.\textwidth]{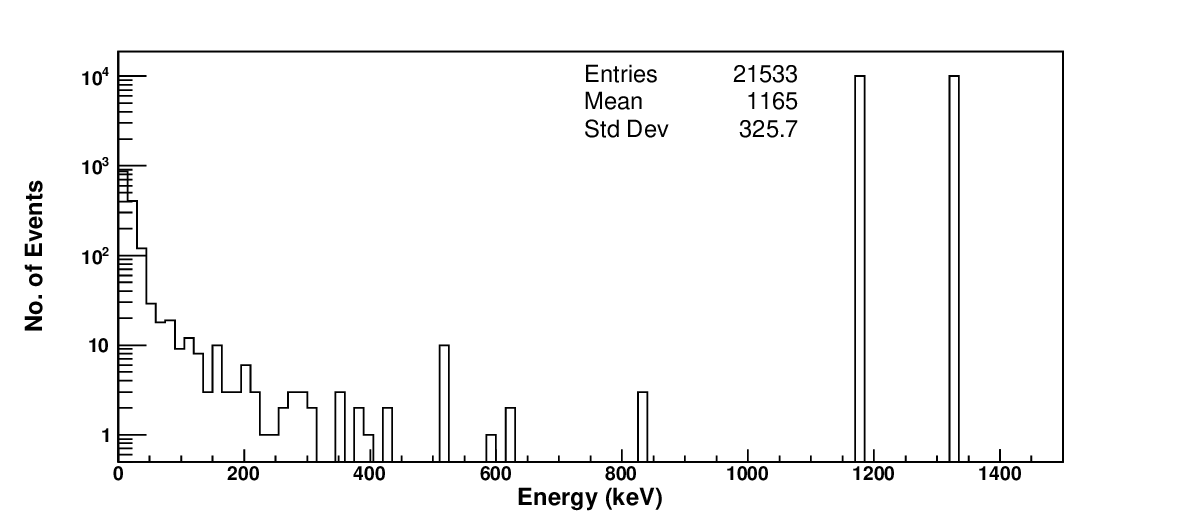}

\caption{The figures show the Cobalt decay in straw tube detectors filled with $Ar:CO_{2}$ (top) and $Xe:CO_{2}$ (bottom) gas mixtures respectively, both in the ratio of 70:30. The two peaks show that the Cobalt decay at 1.1 and 1.3 MeV energy respectively. \label{fig:Two}}
\end{figure}

The detector was irradiated with a gamma source $^{60}$CO for its decay \cite{cobalt}. The physics lists used for $^{60}$CO decay were, GammaNuclearPhysics and G4DecayPhysics. The figure~\ref{fig:One} shows the geant4 display for straw tube detectors irradiated with $^{60}$CO. The $^{60}$CO source passes through graphite and transition radiators and deposit its energy in $Ar:CO_{2}$ and $Xe:CO_{2}$ gas mixture which were taken in the ratio of 70:30. The figure~\ref{fig:Two} shows the $^{60}$CO decay in straw tube detectors filled with $Ar:CO_{2}$ and $Xe:CO_{2}$ gas mixtures respectively taken in the 70:30 ratio. The two peaks at 1.1 and 1.3 MeV show the $^{60}$CO decay into a stable $^{60}$Ni element.

\subsection{XTR Study with Straw Tube Detectors}

The XTR photons are released when a relativistic particle traverses in an in-homogeneous medium i.e., through the boundary between materials of different electrical properties \cite{straw9}--\cite{straw12a}. The relation for transition radiation (TR) production and its dependence on Lorentz factor ($\gamma$) is given by eq.~\eqref{eqxtr}:

\begin{eqnarray}
\label{eqxtr}
\frac{dW}{d\omega} = \frac{4\alpha}{\sigma (\kappa +  1)} (1-exp(-N_{f}\sigma)) \times \Sigma_{n} \theta_{n} \biggl(\frac{1}{\rho_{1} + \theta_{n}} - \frac{1}{\rho_{2} + \theta_{n}}  \biggr)^{2} \biggl[1- cos(\rho_{1} + \theta_{n}) \biggr]
\end{eqnarray}

\noindent Where,
\begin{eqnarray}
\label{eqxtr1}
\rho_{i} = \frac{\omega l_{1}}{2\beta c (\gamma^{-2} + \xi_{i}^{2})}, \nonumber\\
\kappa = \frac{l_{2}}{l_{1}}, \nonumber \\
\theta_{n} = \frac{2\pi n - (\rho_{1} + \kappa\rho_{2})}{1+ \kappa} > 0, \nonumber \\
\gamma = \sqrt (1-(v/c)^{2}) \nonumber
\end{eqnarray}

The eq.~\eqref{eqxtr} shows the TR yield and energy spectrum dependence on various factors like Lorentz factor, number of foils, foil thickness and spacing. In the equation, W is energy spectrum radiated by a charged particle. Also, $\alpha$ = 1/137 is the fine structure constant, $\sigma$ is the absorption cross section for the foil and gas. $N_{f}$ represents the number of foils, $\omega$ is plasma frequency for the two media, $\xi_{i}$ <<1, $l_{1}$ is foil thickness, $l_{2}$ is the spacing between the foils, $\theta$ is emission angle. $\beta$ = $v/c$, v is the velocity of particle and c is speed of light. The XTR photon yield gives the amount of XTR photon emitted from the transition of particles (electrons) in different media.

\begin{figure}[htbp]
\centering
\includegraphics[width=1.\textwidth]{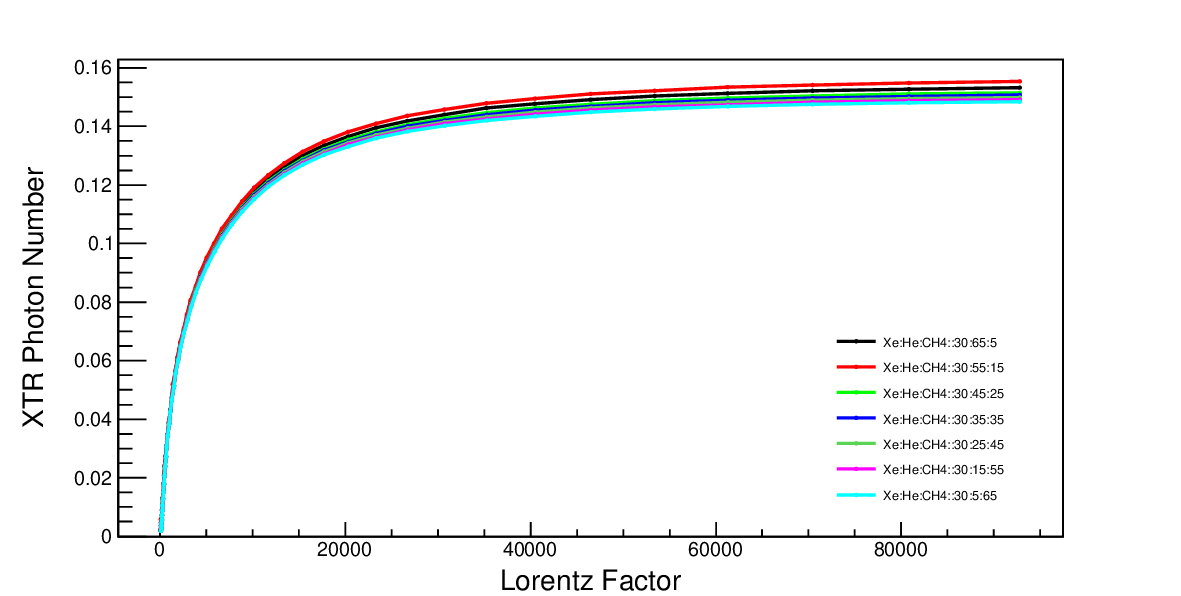}

\caption{XTR photon number as a function of the Lorentz factor for different ratio of Xe:He:$CH_{4}$ gas mixture.\label{fig:ThreeOne}}
\end{figure}

The geant4 G4StrawTubeXTRadiator model was chosen for the XTR study. In order to use the xenon-based gas mixtures for the study of x-ray transition radiation first step done was to optimize $XeHeCH_{4}$ gas mixture by keeping xenon ratio constant i.e., 30\% and changing the ratio of $He:CH_{4}$ gas mixture as mentioned in the table~\ref{tableL}. Figure~\ref{fig:ThreeOne} shows the XTR photon yield as a function of the Lorentz factor for different compositions of $XeHeCH_{4}$ gas mixture. The figure shows that as the Lorentz factor increases the XTR yield also increases and gets saturated at the Lorentz factor of 46520 as the XTR yield saturates with the multiple foils \cite{straw12b}. Among all the composition of $XeHeCH_{4}$ gas mixture, the ratio 30:55:15 shows the highest XTR yield. 
\begin{table}[htbp]
\centering
\caption{The optimization of $XeHeCH_{4}$ gas mixture by changing the $He:CH_{4}$ ratio.\label{tableL}}
%\smallskip
\begin{tabular}{|c|c|}
\hline
Gases & Different composition of He gas.\\
\hline
$XeHeCH_{4}$ & 30:65:5 \\
$XeHeCH_{4}$ & 30:55:15 \\
$XeHeCH_{4}$ & 30:45:25\\
$XeHeCH_{4}$ & 30:35:35\\
$XeHeCH_{4}$ & 30:25:45 \\
$XeHeCH_{4}$ & 30:15:55 \\
$XeHeCH_{4}$ & 30:5:65 \\

\hline
\end{tabular}
\end{table}

\begin{figure*}[htbp]
\centering
\includegraphics[width=0.9\textwidth]{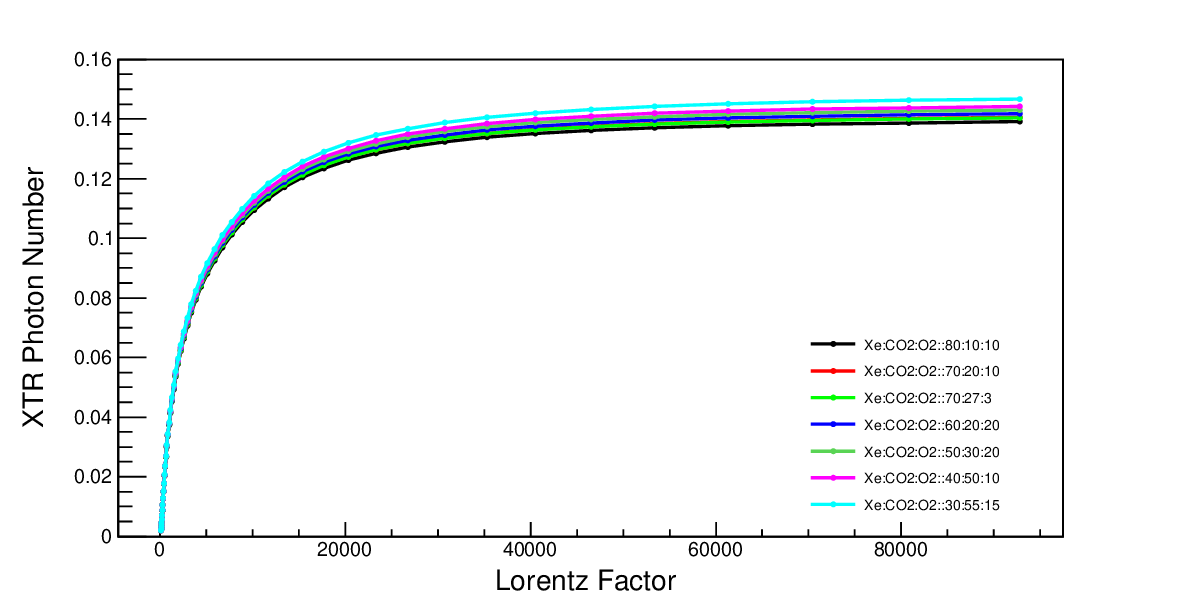}
\qquad
\includegraphics[width=0.9\textwidth]{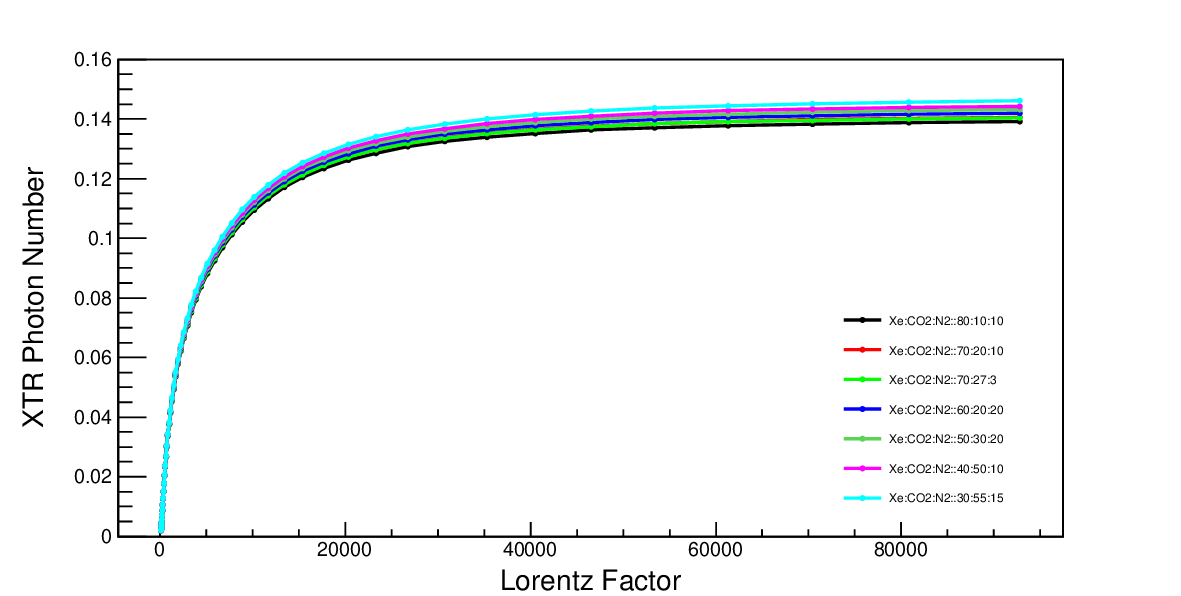}
\qquad
\includegraphics[width=0.9\textwidth]{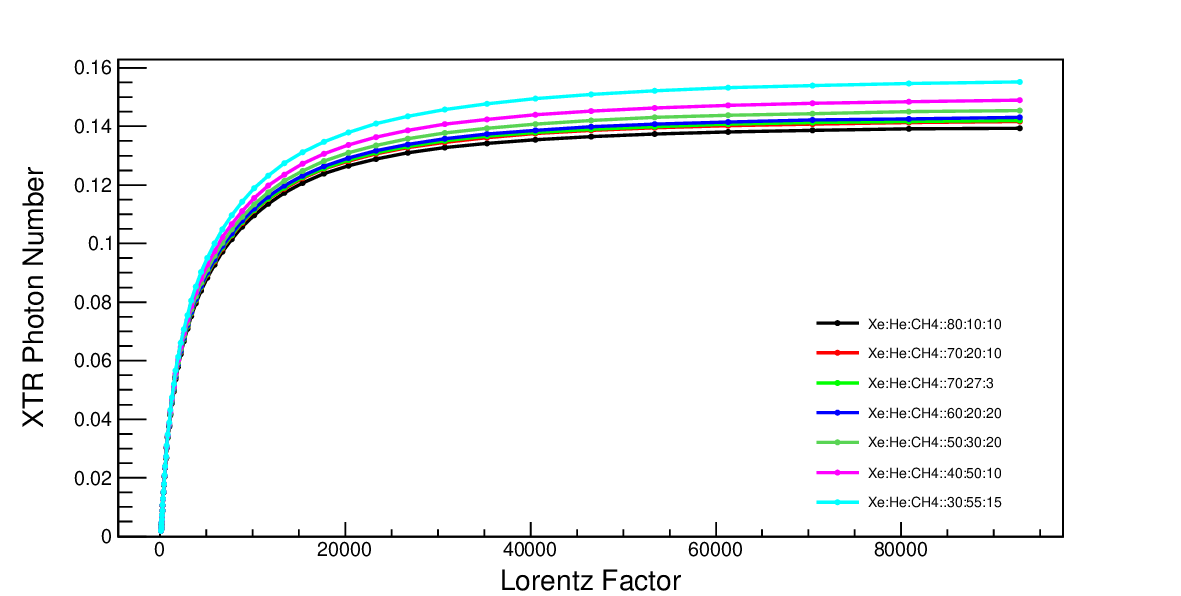}\\
\caption{XTR photon number as a function of the Lorentz factor for different gas mixtures i.e., $XeCO_{2}O_{2}$, $XeCO_{2}N_{2}$, $XeHeCH_{4}$ respectively, taken in the different ratios as mentioned in table~\ref{table1}. \label{fig:ThreeTwo}}
\end{figure*}

\begin{figure*}[htbp]
\centering
\includegraphics[width=0.9\textwidth]{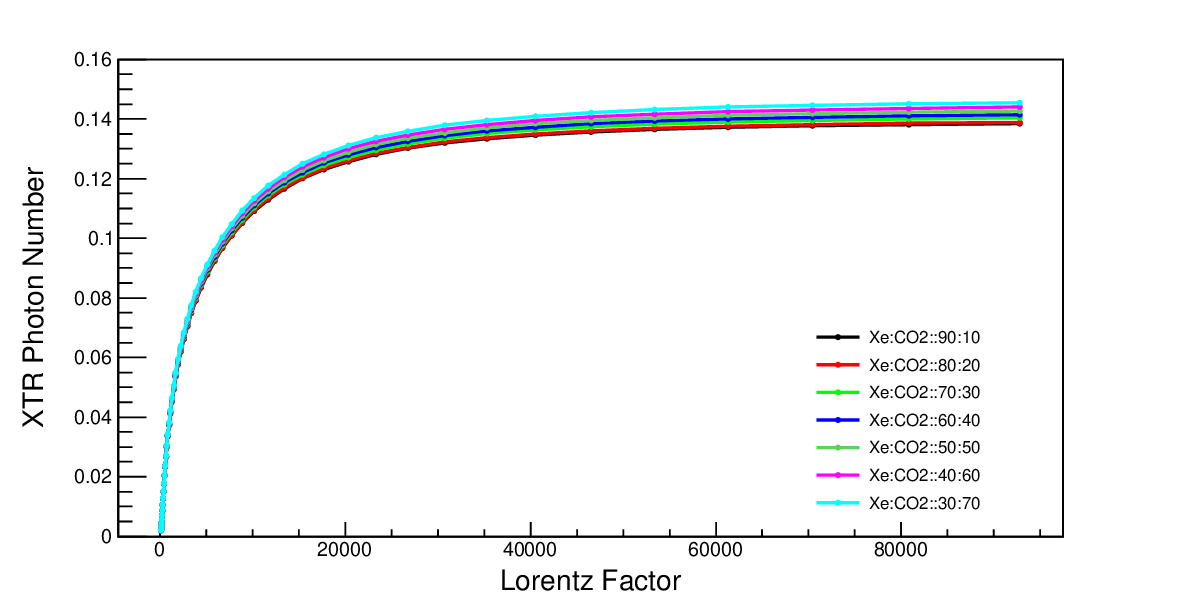}
\qquad
\includegraphics[width=0.9\textwidth]{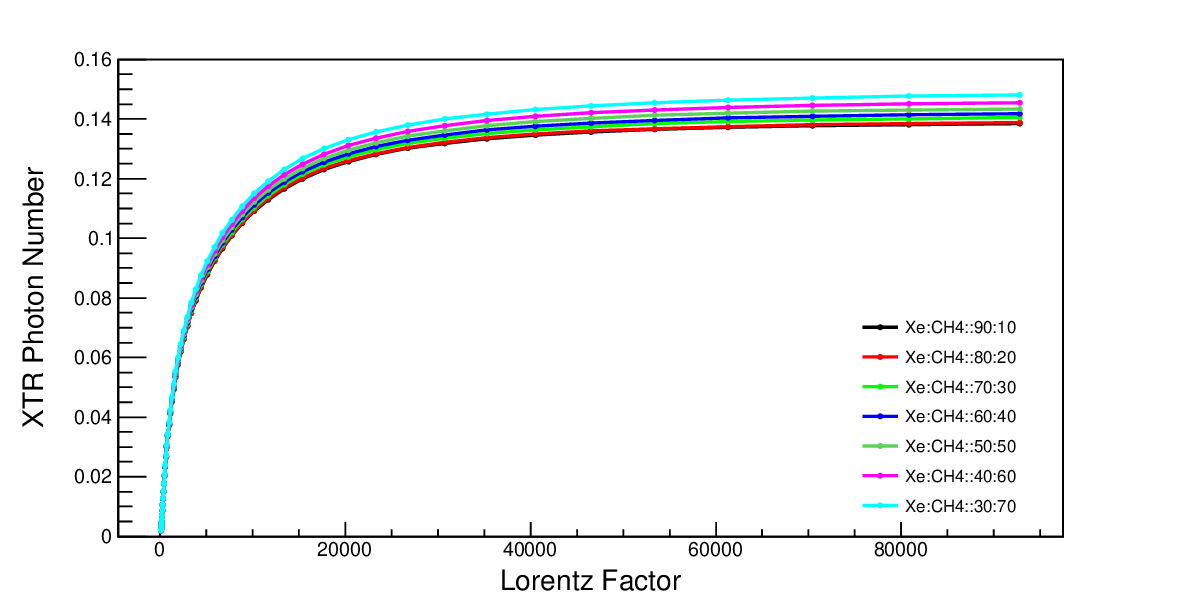}
\qquad
\includegraphics[width=0.9\textwidth]{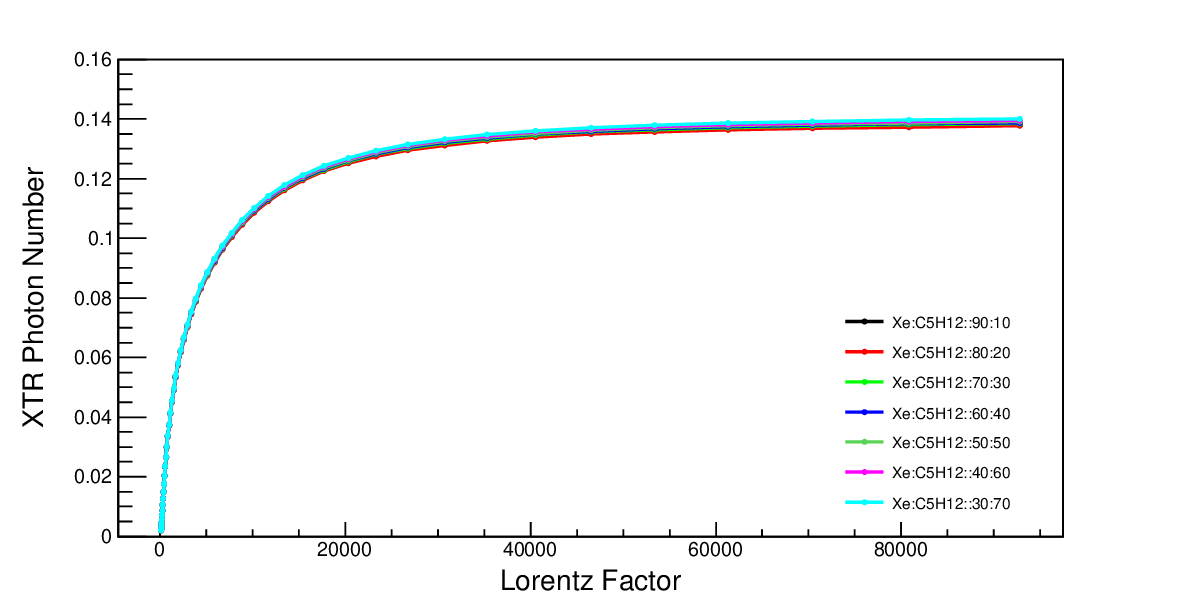}

\caption{XTR photon number as a function of the Lorentz factor for different gas mixtures i.e., $XeCO_{2}$, $XeCH_{4}$, $XeC_{5}H_{12}$ respectively, taken in the different ratios as mentioned in table~\ref{table1}. \label{fig:ThreeThree}}
\end{figure*}

\begin{figure*}[htbp]
\centering
\includegraphics[width=1.\textwidth]{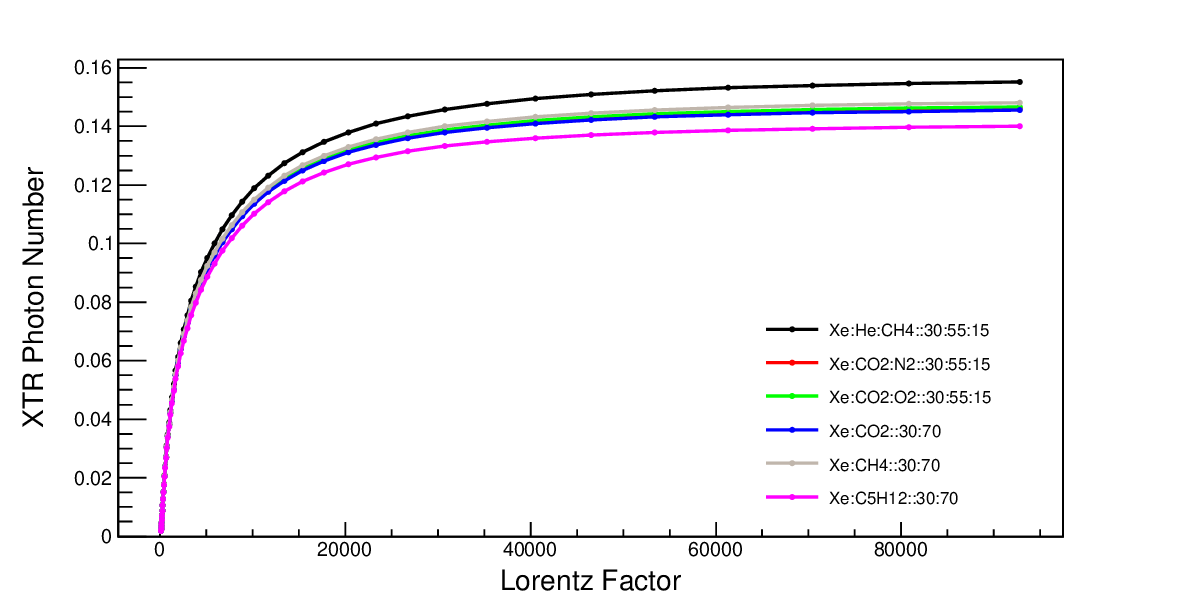}

\caption{XTR photon number as a function of the Lorentz factor for different gas mixtures taken in the ratio of 30:55:15 and 30:70 for three and two gas mixtures respectively.\label{fig:ThreeFour}}
\end{figure*}

%\clearpage
The optimization of gas mixtures has been continued in order to select the gas mixture that shows maximum transition radiation. The different gas mixtures as mentioned in table~\ref{table1} have been used for the study. These gas mixtures have been optimized by changing the ratios of different gas compositions. Figures~\ref{fig:ThreeTwo} and~\ref{fig:ThreeThree} show the optimization of the gas mixtures i.e., $XeCO_{2}O_{2}$, $XeCO_{2}N_{2}$, $XeHeCH_{4}$, $XeCO_{2}$, $XeCH_{4}$, $XeC_{5}H_{12}$ for the study of transition radiation. Out of all the compositions the ratios 30:55:15 and 30:70 show the maximum transition radiation for three and two gas mixtures respectively. Figure~\ref{fig:ThreeFour} shows the XTR yield as a function of Lorentz factor when the ratios were taken as 30:55:15 and 30:70 for three and two gas mixtures respectively. Among them $Xe:He:CH_{4}$ :: 30:55:15 shows the highest XTR photon yield. The reason behind the highest yield is attributed due to several factors. The helium and methane ($CH_{4}$) gases have relatively low x-ray absorption. The presence of $CO_{2}$ in other xenon gas mixture leads to loss of XTR photons before detection and decreases its yield. Hence the overall performance of $XeHeCH_{4}$ gas mixture for the production of XTR is higher than all the other gas mixtures.

\subsection{Study of Primary Ionization with Straw Tube Detectors}
\label{PrimaryI}

\begin{figure}[htbp]
\centering
\includegraphics[width=1.\textwidth]{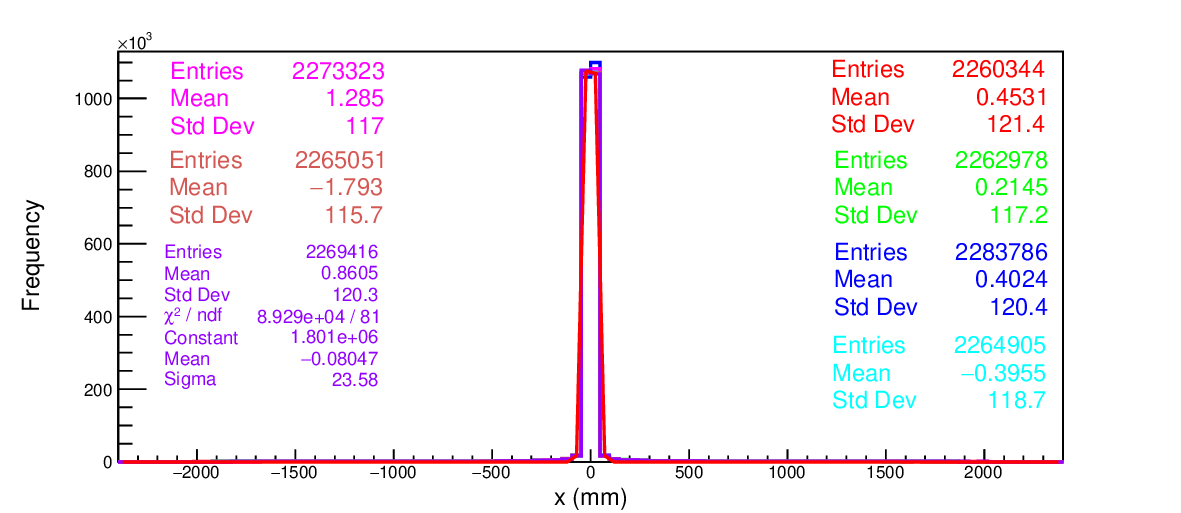}
\qquad
\includegraphics[width=1.\textwidth]{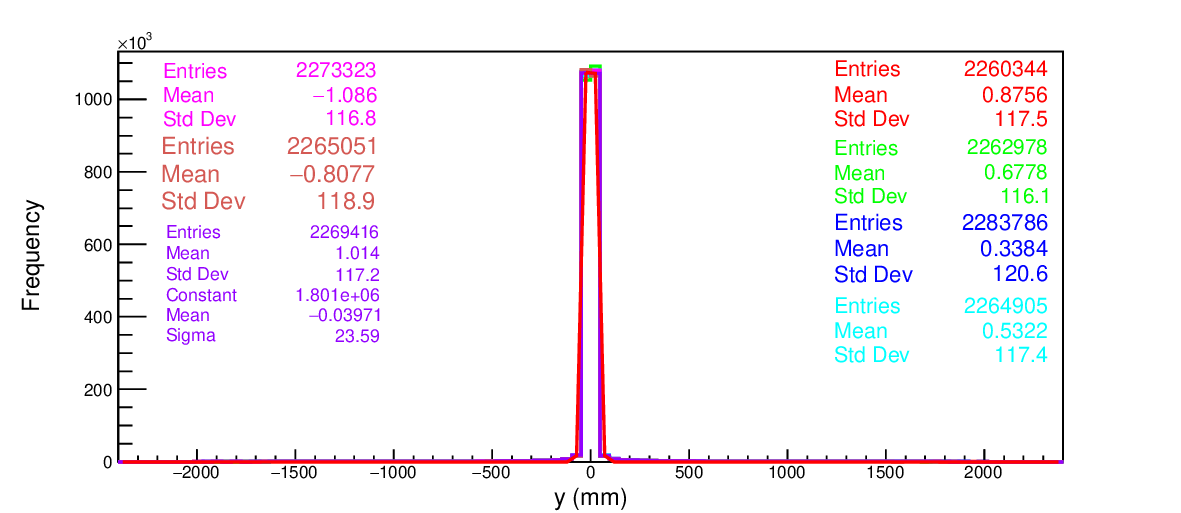}\\
\qquad
\includegraphics[width=1.\textwidth]{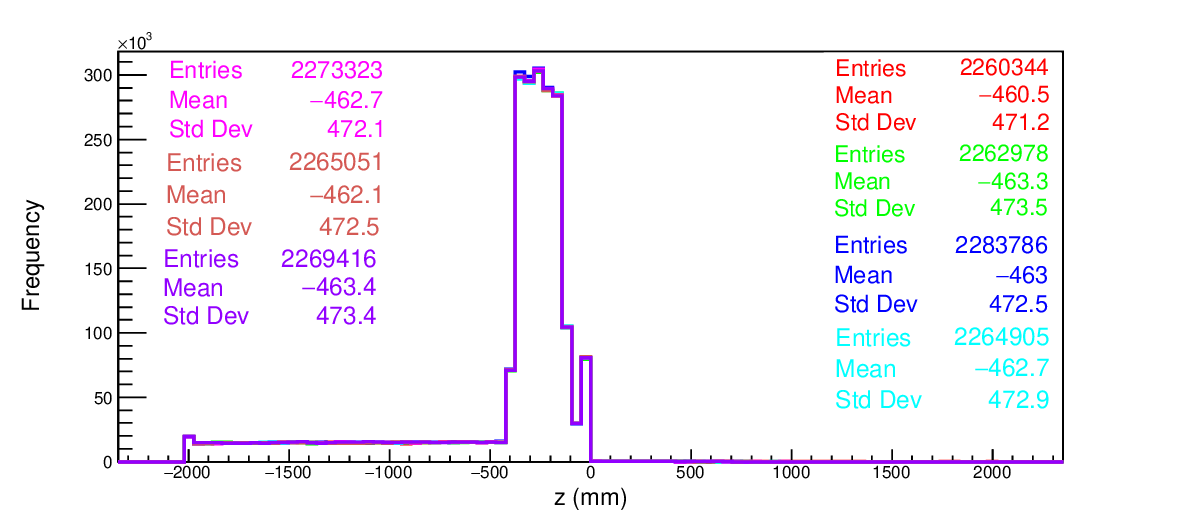}
\caption{The x-, y- and z- distributions of the primary ionizations when electrons of 1 GeV were shot in $XeCO_{2}$ gas mixture as mentioned in table~\ref{table1}. Note that, the colour scheme for the gas mixtures are: red - 90:10, green - 80:20, blue - 70:30, cyan - 60:40, pink - 50:50, brown - 40:60, purple - 30:70. \label{fig:FiveOne}}
\end{figure}

\begin{figure}[htbp]
\centering
\includegraphics[width=1.\textwidth]{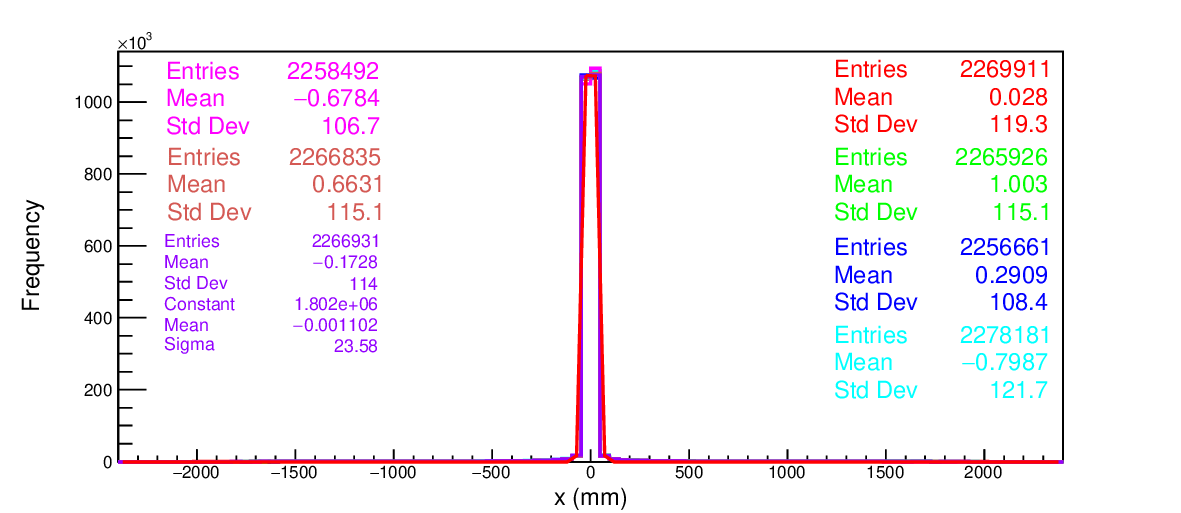}
\qquad
\includegraphics[width=1.\textwidth]{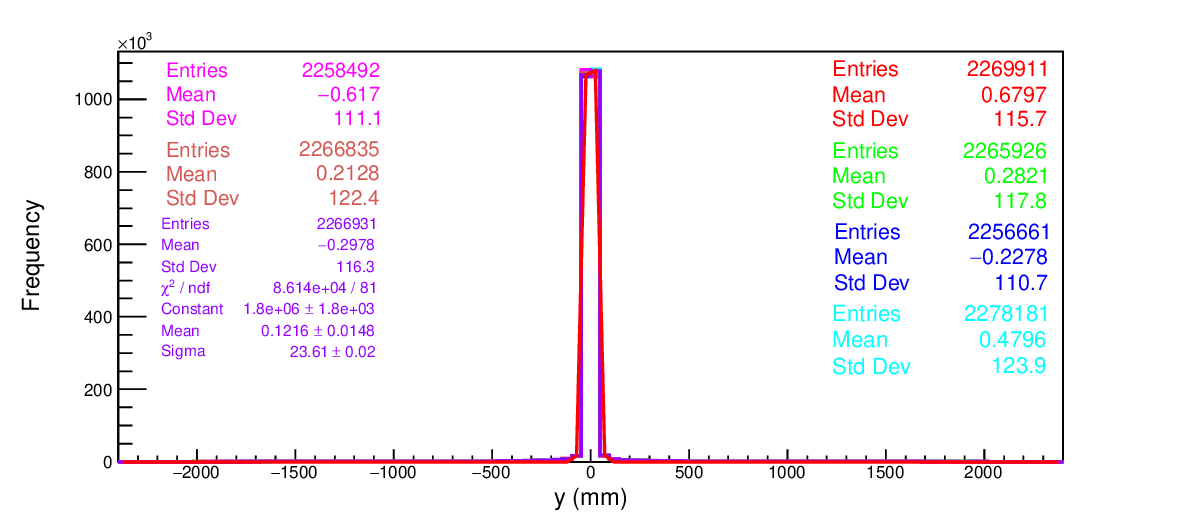}\\
\qquad
\includegraphics[width=1.\textwidth]{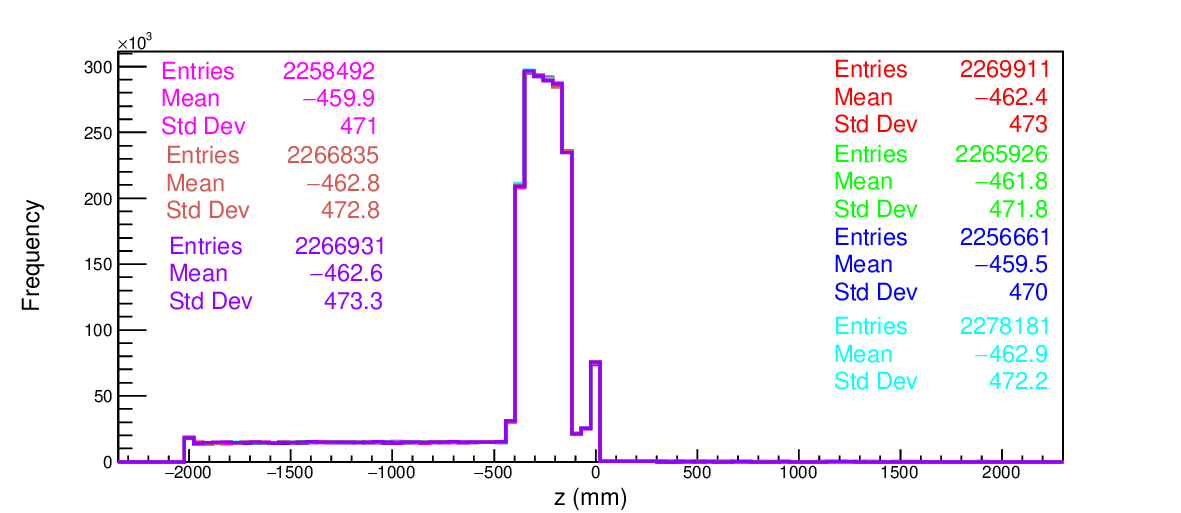}
\caption{The x-, y- and z- distributions of the primary ionizations when electrons of 1 GeV were shot in $XeCO_{2}N_{2}$ gas mixture as mentioned in table~\ref{table1}. Note that, the colour scheme for the gas mixtures are: red - 70:27:3, green - 80:10:10, blue - 70:20:10, cyan - 60:20:20, pink - 50:30:20, brown - 40:50:10, purple - 30:55:15.\label{fig:FiveTwo}}
\end{figure}

\begin{figure}[htbp]
\centering
\includegraphics[width=1.\textwidth]{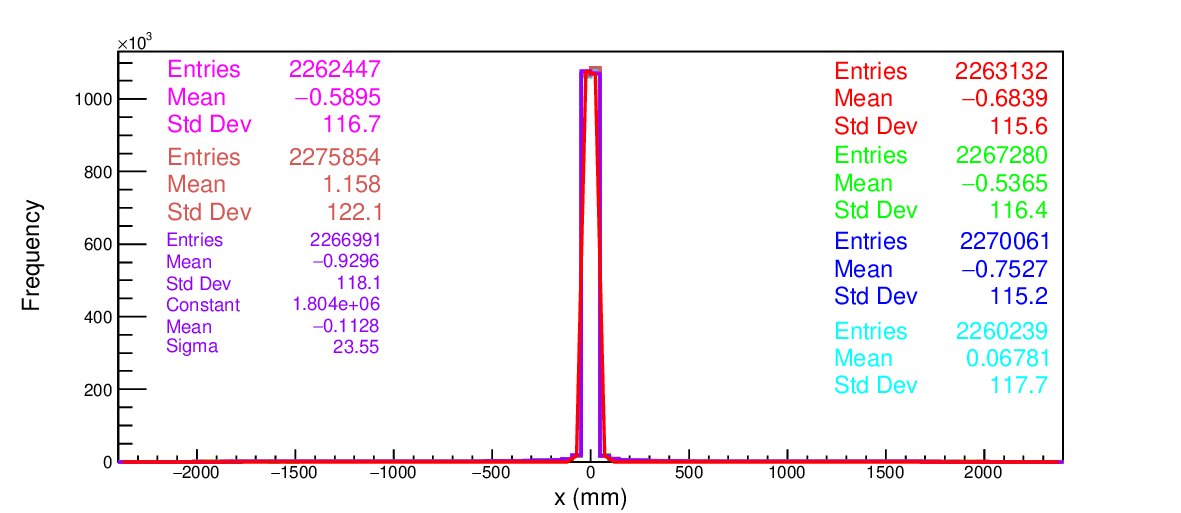}
\qquad
\includegraphics[width=1.\textwidth]{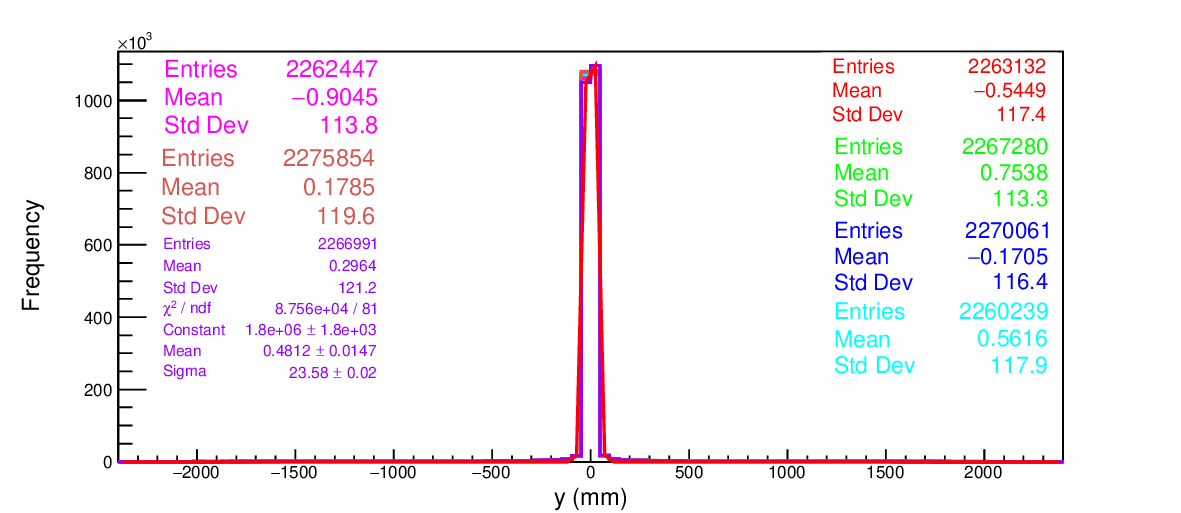}\\
\qquad
\includegraphics[width=1.\textwidth]{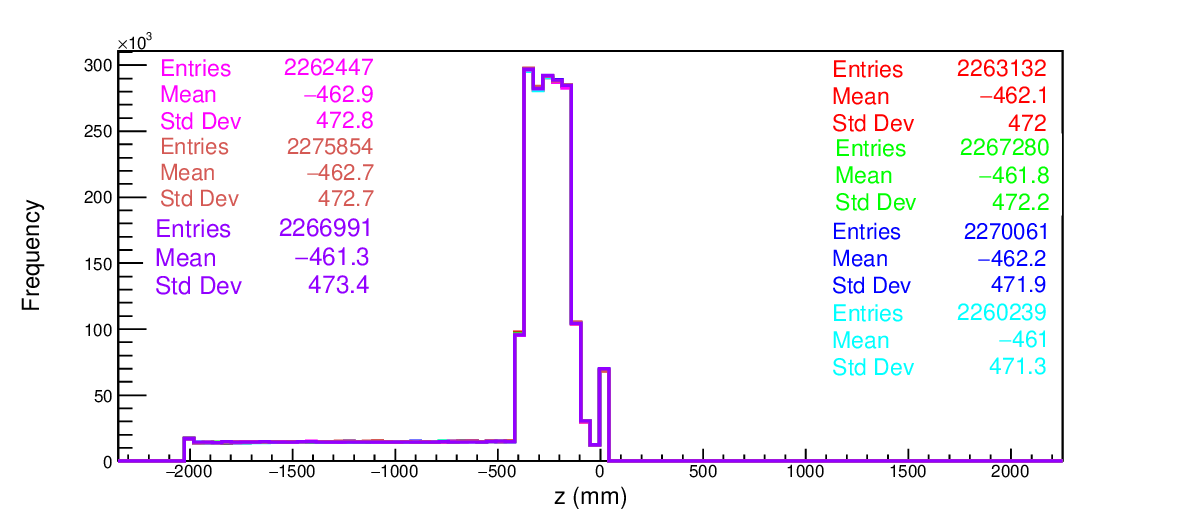}
\caption{The x-, y- and z- distributions of the primary ionizations when electrons of 1 GeV were shot in $XeHeCH_{4}$ gas mixture as mentioned in table~\ref{table1}. Note that, the colour scheme for the gas mixtures are: red - 70:27:3, green - 80:10:10, blue - 70:20:10, cyan - 60:20:20, pink - 50:30:20, brown - 40:50:10, purple - 30:55:15.\label{fig:FiveThree}}
\end{figure}

\begin{figure}[htbp]
\centering
\includegraphics[width=1.\textwidth]{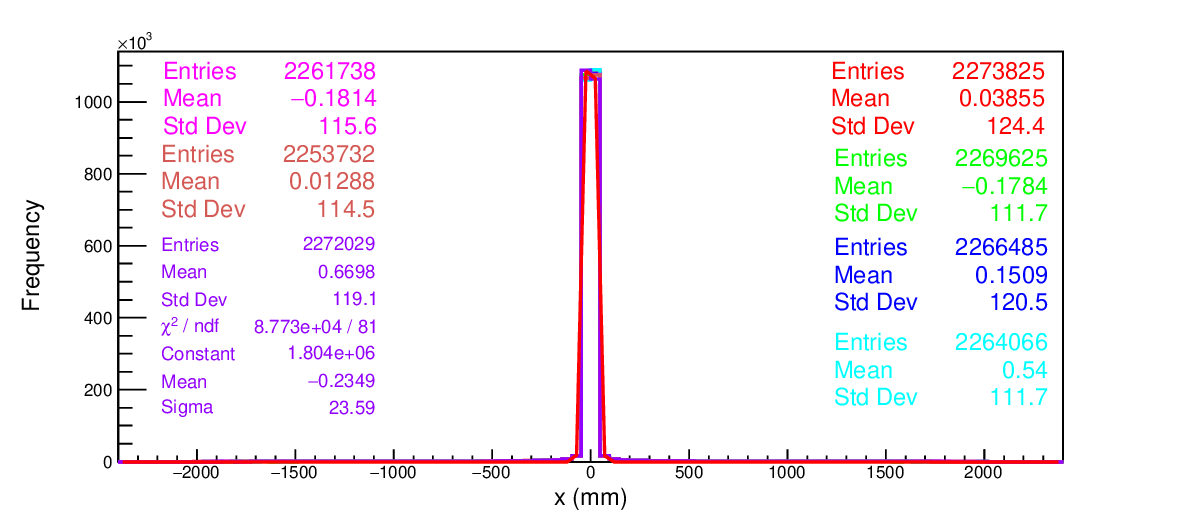}
\qquad
\includegraphics[width=1.\textwidth]{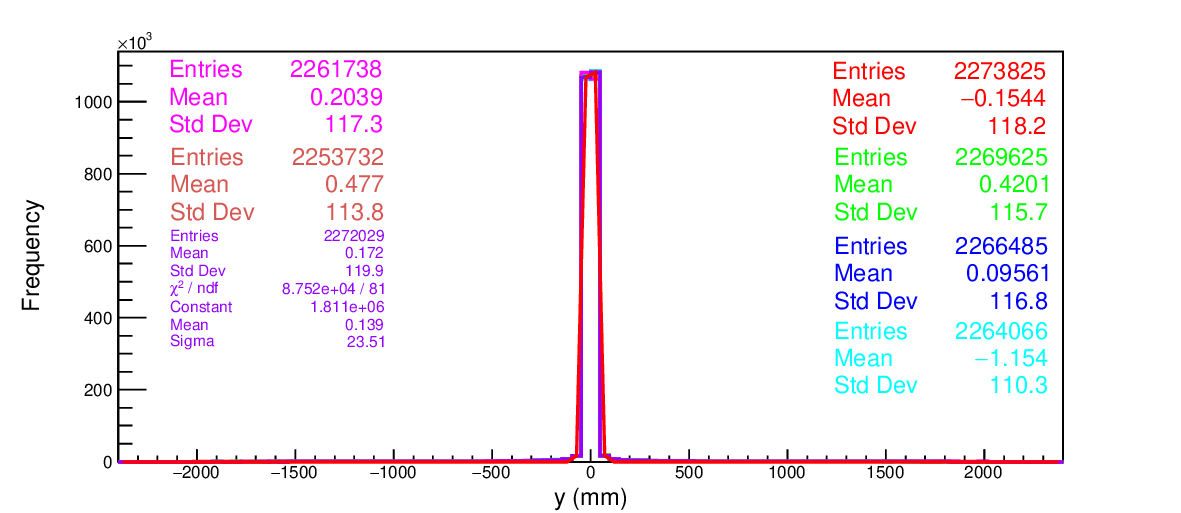}\\
\qquad
\includegraphics[width=1.\textwidth]{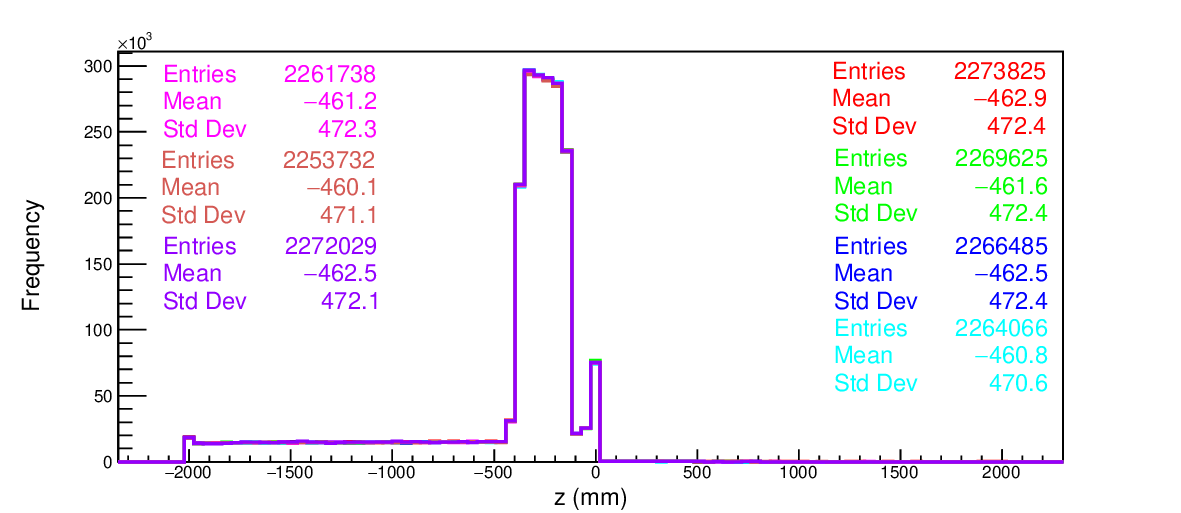}
\caption{The x-, y- and z- distributions of the primary ionizations when electrons of 1 GeV were shot in $XeCO_{2}O_{2}$ gas mixture as mentioned in table~\ref{table1}. Note that, the colour scheme for the gas mixtures are: red - 70:27:3, green - 80:10:10, blue - 70:20:10, cyan - 60:20:20, pink - 50:30:20, brown - 40:50:10, purple - 30:55:15. \label{fig:FiveFour}}
\end{figure}

\begin{figure}[htbp]
\centering
\includegraphics[width=1.\textwidth]{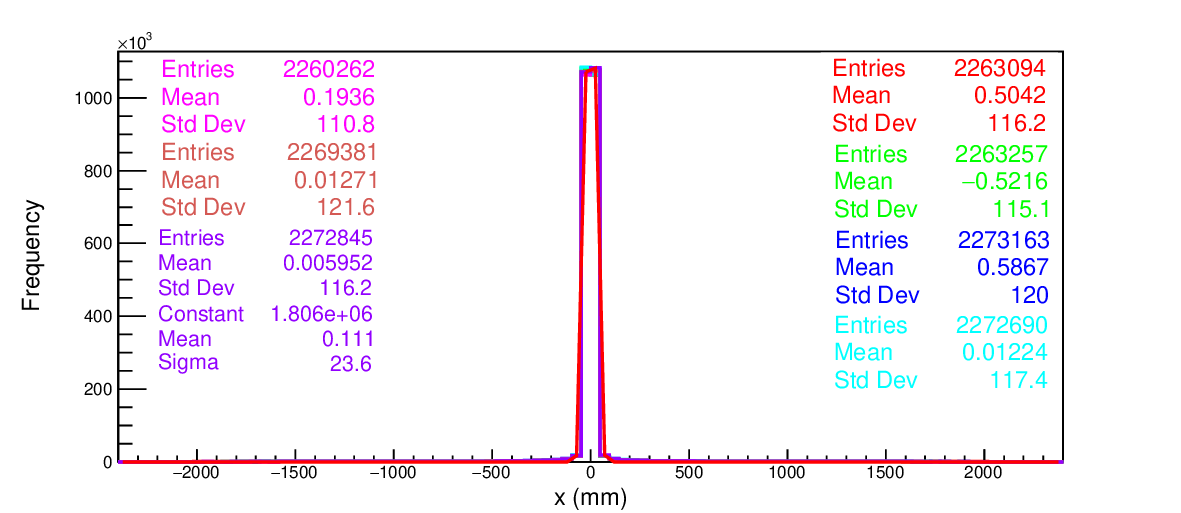}
\qquad
\includegraphics[width=1.\textwidth]{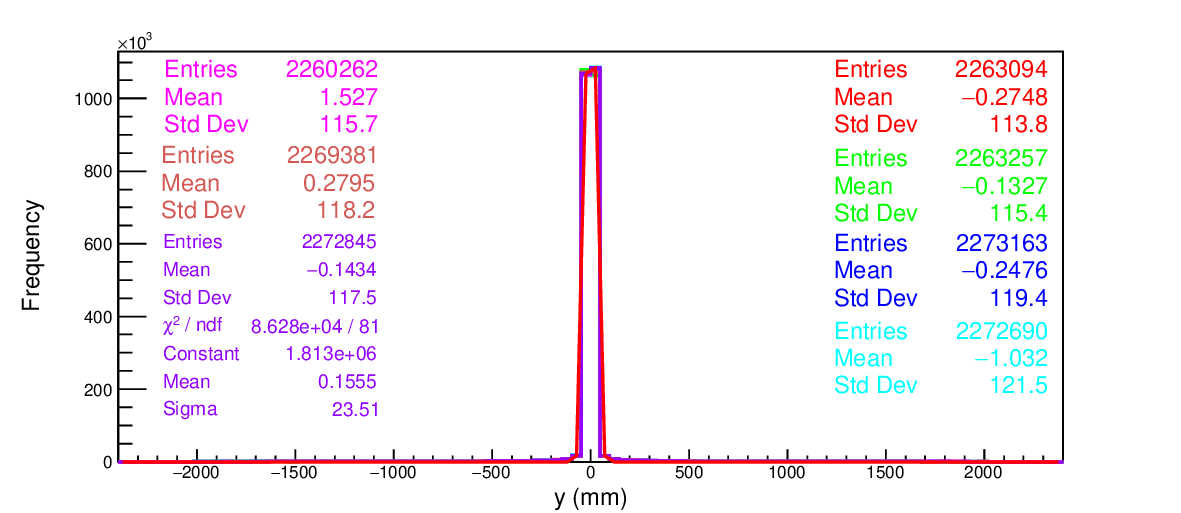}\\
\qquad
\includegraphics[width=1.\textwidth]{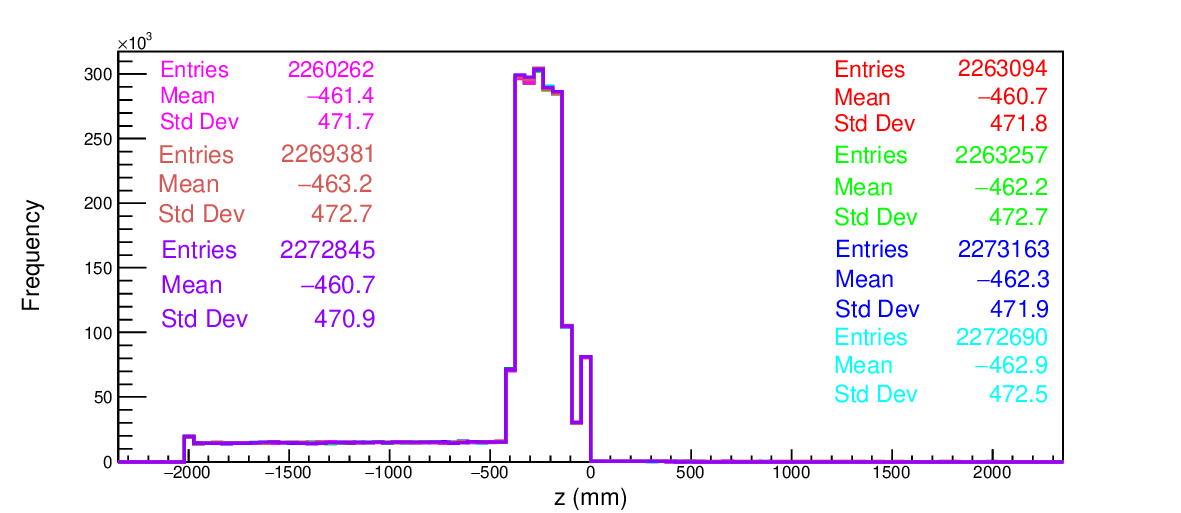}
\caption{The x-, y- and z- distributions of the primary ionizations when electrons of 1 GeV were shot in $XeC_{5}H_{12}$ gas mixture as mentioned in table~\ref{table1}. Note that, the colour scheme for the gas mixtures are: red - 90:10, green - 80:20, blue - 70:30, cyan - 60:40, pink - 50:50, brown - 40:60, purple - 30:70.\label{fig:FiveFive}}
\end{figure}

\begin{figure}[htbp]
\centering
\includegraphics[width=1.\textwidth]{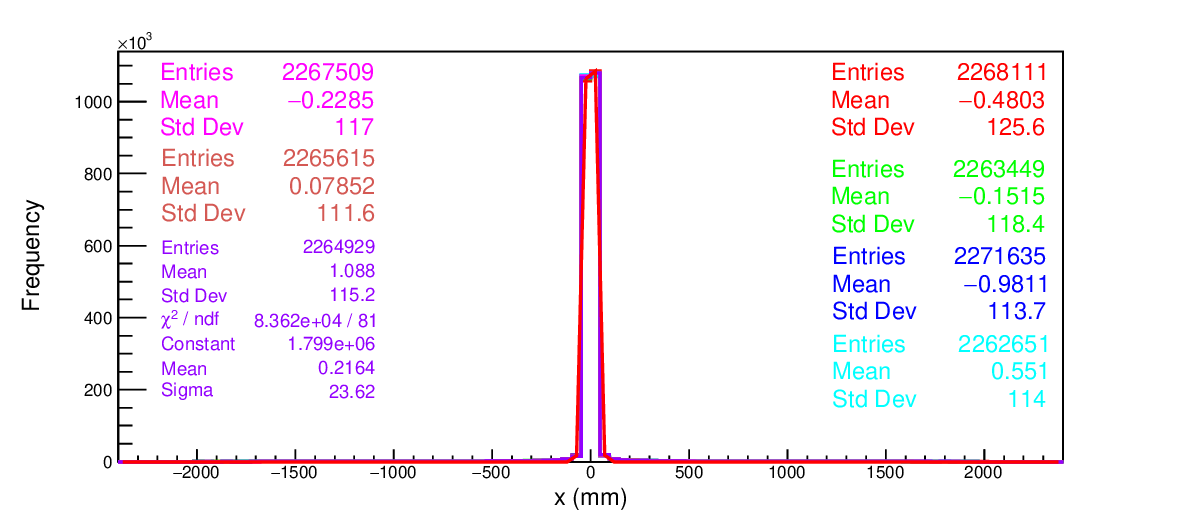}
\qquad
\includegraphics[width=1.\textwidth]{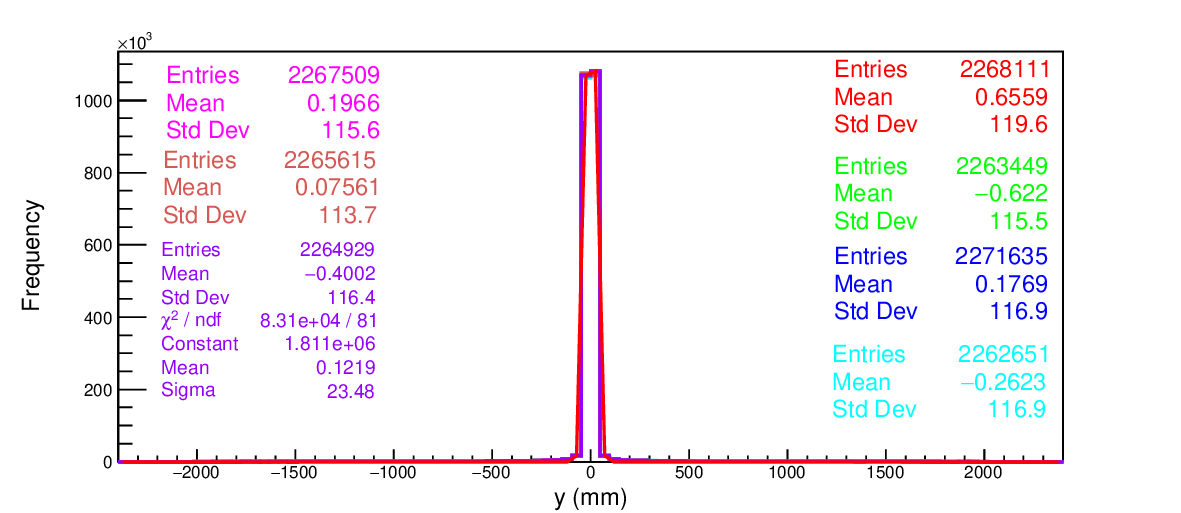}\\
\qquad
\includegraphics[width=1.\textwidth]{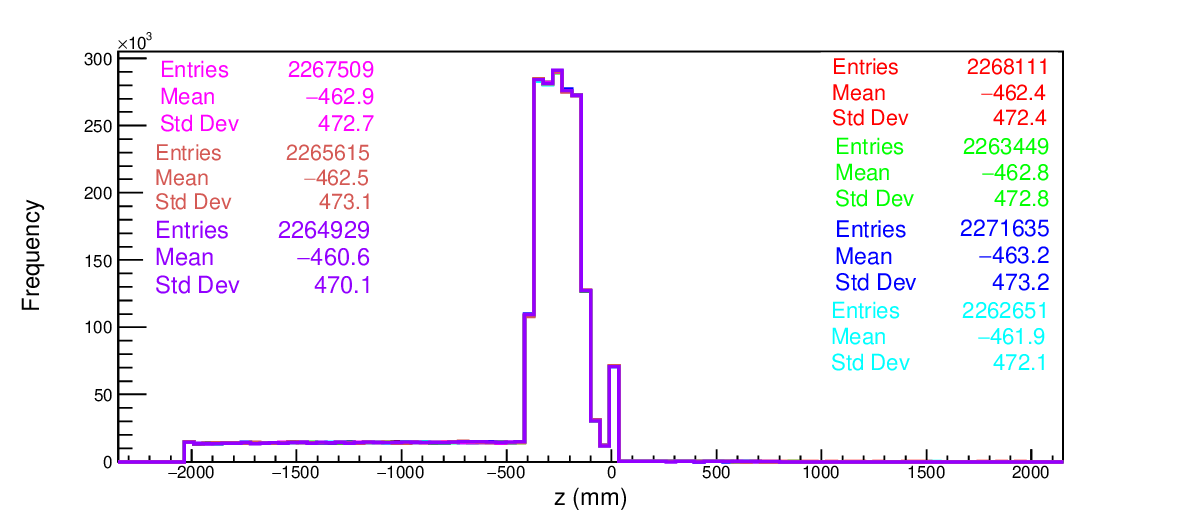}
\caption{ The x-, y- and z- distributions of the primary ionizations when electrons of 1 GeV were shot in $XeCH_{4}$ gas mixture as mentioned in table~\ref{table1}. Note that, the colour scheme for the gas mixtures are: red - 90:10, green - 80:20, blue - 70:30, cyan - 60:40, pink - 50:50, brown - 40:60, purple - 30:70.\label{fig:FiveSix}}
\end{figure}

\begin{figure}[htbp]
\centering
\includegraphics[width=1.\textwidth]{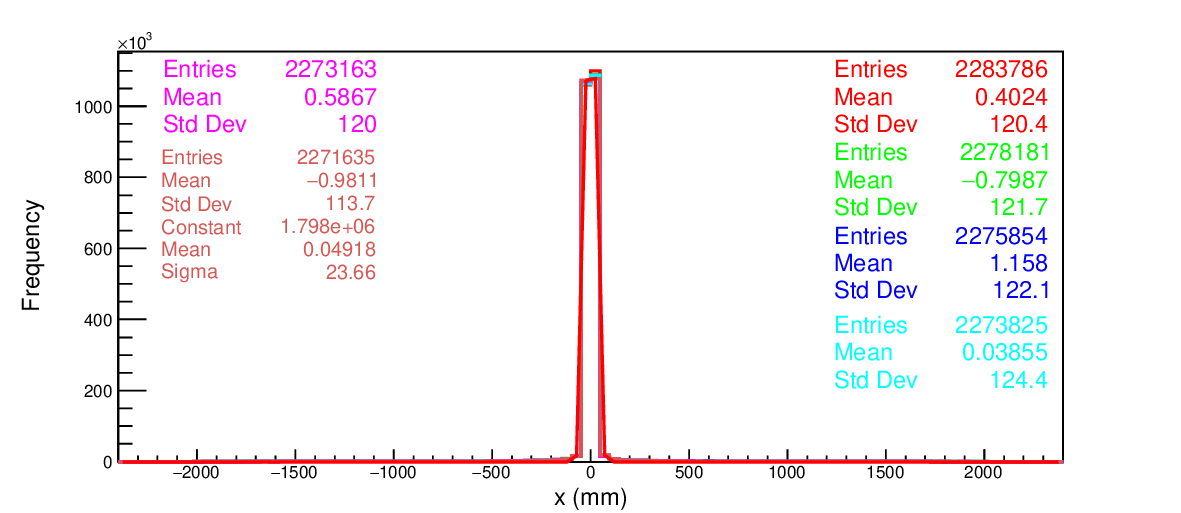}
\qquad
\includegraphics[width=1.\textwidth]{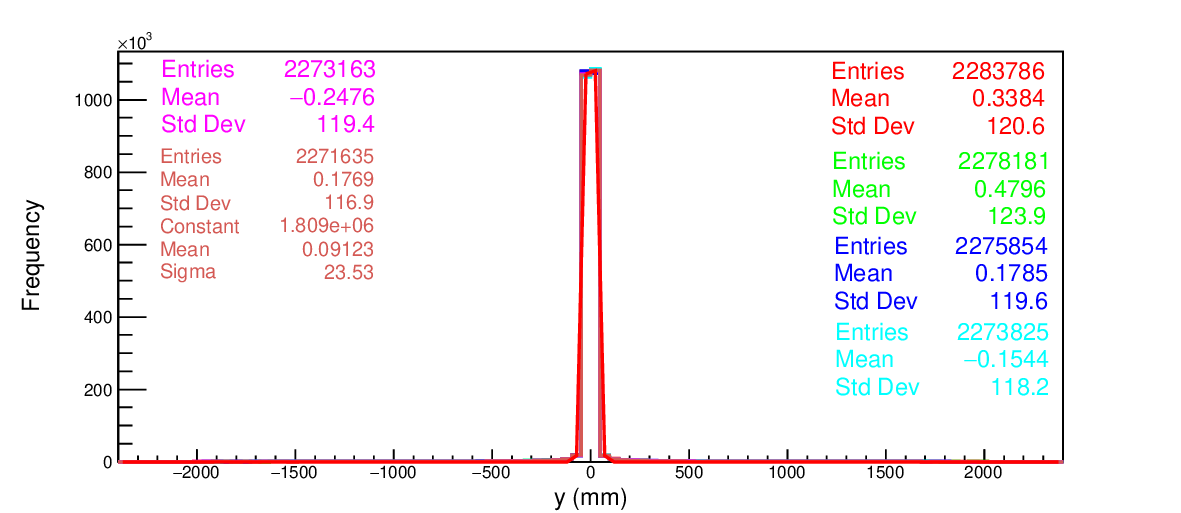}\\
\qquad
\includegraphics[width=1.\textwidth]{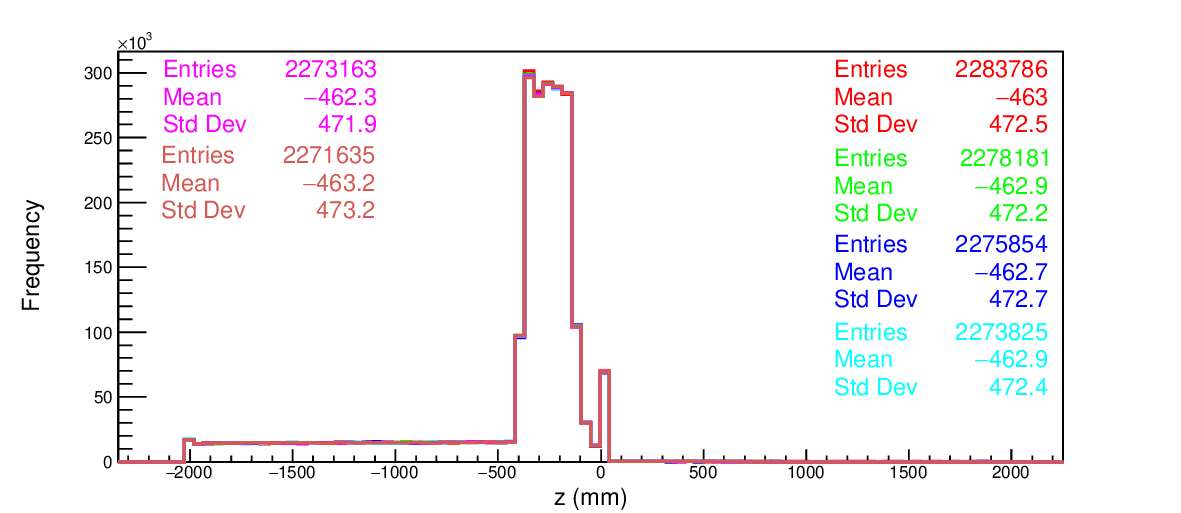}
\caption{The x-, y- and z- distributions of the primary ionizations when electrons of 1 GeV were shot in selected gas mixtures as mentioned in table~\ref{table21}. Note that, the colour scheme for the gas mixtures are: red - $XeCO_{2}$ :: 70:30, green - $XeCO_{2}N_{2}$ :: 60:20:20, blue - $XeHeCH_{4}$ :: 40:50:10, cyan - $XeCO_{2}O_{2}$ :: 70:27:3, pink - $XeC_{5}H_{12}$ :: 70:30, brown - $XeCH_{4}$ :: 70:30. \label{fig:FiveSeven}}
\end{figure}

\begin{figure}[htbp]
\centering
\includegraphics[width=1.\textwidth]{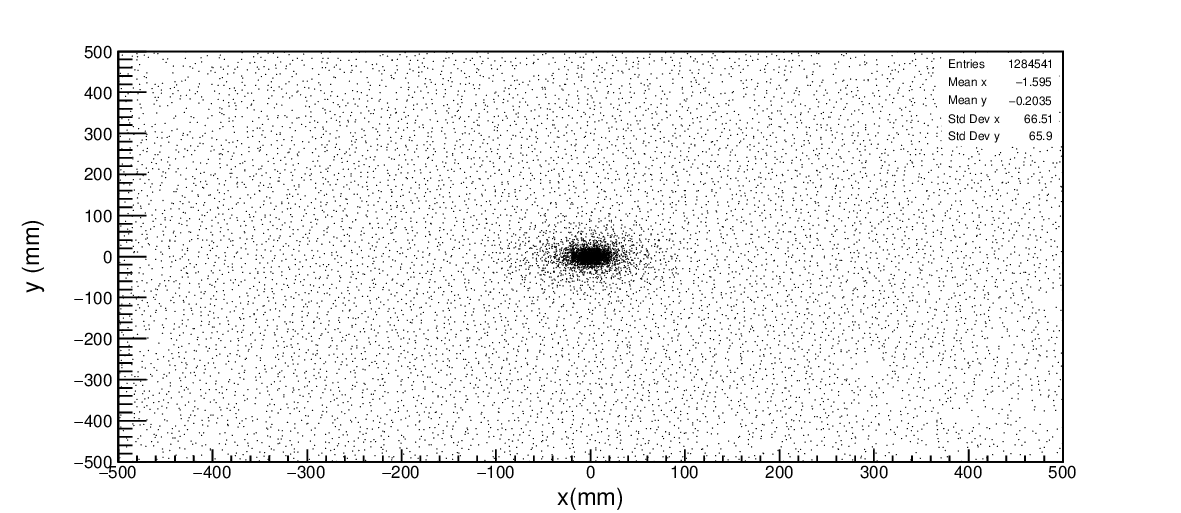}

\caption{The figure shows the variation of x- and y- coordinates in 2 dimensions in $XeCO_{2}$ :: 70:30 gas mixture.\label{fig:Six}}
\end{figure}

The different gas mixtures as mentioned in table~\ref{table1} have been analyzed to obtain primaries and the spatial co-ordinate distributions \cite{primary1}. 10,000 electrons of 1 GeV were shot from x=0 cm, y=0 cm, z=0 cm in the six gas mixtures as discussed in the table~\ref{table1}. The electrons were shot along the z-direction. The physics list chosen for the interactions of particles with the gas mixtures was G4EmStandardPhysics. The figures~\ref{fig:FiveOne}, ~\ref{fig:FiveTwo}, ~\ref{fig:FiveThree}, ~\ref{fig:FiveFour}, ~\ref{fig:FiveFive}, and~\ref{fig:FiveSix} show the x-, y- and z- distributions of electrons with different gas mixtures i.e., $XeCO_{2}$, $XeCO_{2}N_{2}$, $XeHeCH_{4}$, $XeCO_{2}O_{2}$, $XeC_{5}H_{12}$, $XeCH_{4}$ respectively. The optimization of the primary ionization using the various gas mixtures has been done. The selection of gas mixture has been done by changing the ratios of individual gas compositions as given in the table~\ref{table1}. The figures also show that the x-, y- distributions are Gaussian as have been fit with Gaussian function. The x- and y- distributions show the Gaussian distributions due to large statistics \cite{primary1}-\cite{primary2}. The electrons were shot along negative z direction. The figures show the z- distribution started from -2000 mm until it was stopped in the gas medium. As observed from the figure~\ref{fig:FiveOne} from x-, y-, z-distributions the primary ionization for $XeCO_{2}$::70:30 is 2283786 which is highest among all the compositions of $XeCO_{2}$. Similarly, other gas mixtures were also analyzed from figures~\ref{fig:FiveTwo}--~\ref{fig:FiveSix} in the same way. Hence the table~\ref{table21} was obtained from the analyzed gas mixtures (from figures~\ref{fig:FiveOne}--~\ref{fig:FiveSix}) that were found to show highest primary ionization. From figures~\ref{fig:FiveOne}--~\ref{fig:FiveSix}, those gas mixtures that show maximum primary ionization were considered as ``selected'' gas mixtures. The figure~\ref{fig:FiveSeven} shows the combined x-, y- and z- for the ``selected'' gas mixtures to obtain the highest primary ionization among all the xenon-based gas mixtures. From the figure~\ref{fig:FiveSeven} it has been observed that $XeCO_{2}$ :: 70:30 gas mixture show the highest primary ionization among all the xenon-based gas mixtures. The figure~\ref{fig:Six} shows the view of x- and y- distributions in 2-dimensions in $XeCO_{2}$ :: 70:30 gas mixture. The central part shows that the particles were shot from x=0 cm, y=0 cm.

\begin{table}[htbp]
\centering
\caption{The number of primaries obtained from ``selected'' gas mixtures.\label{table21}}
%\smallskip
\begin{tabular}{|c|c|c|}
\hline
Gases & Composition & No. of primaries\\
\hline
$XeCO_{2}$ & 70:30 & 2283786\\
$XeCO_{2}N_{2}$ & 60:20:20 & 2278181\\
$XeHeCH_{4}$ & 40:50:10 & 2275854\\
$XeCO_{2}O_{2}$ & 70:27:3 & 2273825\\
$XeC_{5}H_{12}$ & 70:30 & 2273163\\
$XeCH_{4}$ & 70:30 & 2271635\\
\hline
\end{tabular}
\end{table}

\subsection{Gas Properties with Straw Tube Detectors}
\label{gasProp}

\begin{figure}[htbp]
\centering
\includegraphics[width=1.\textwidth]{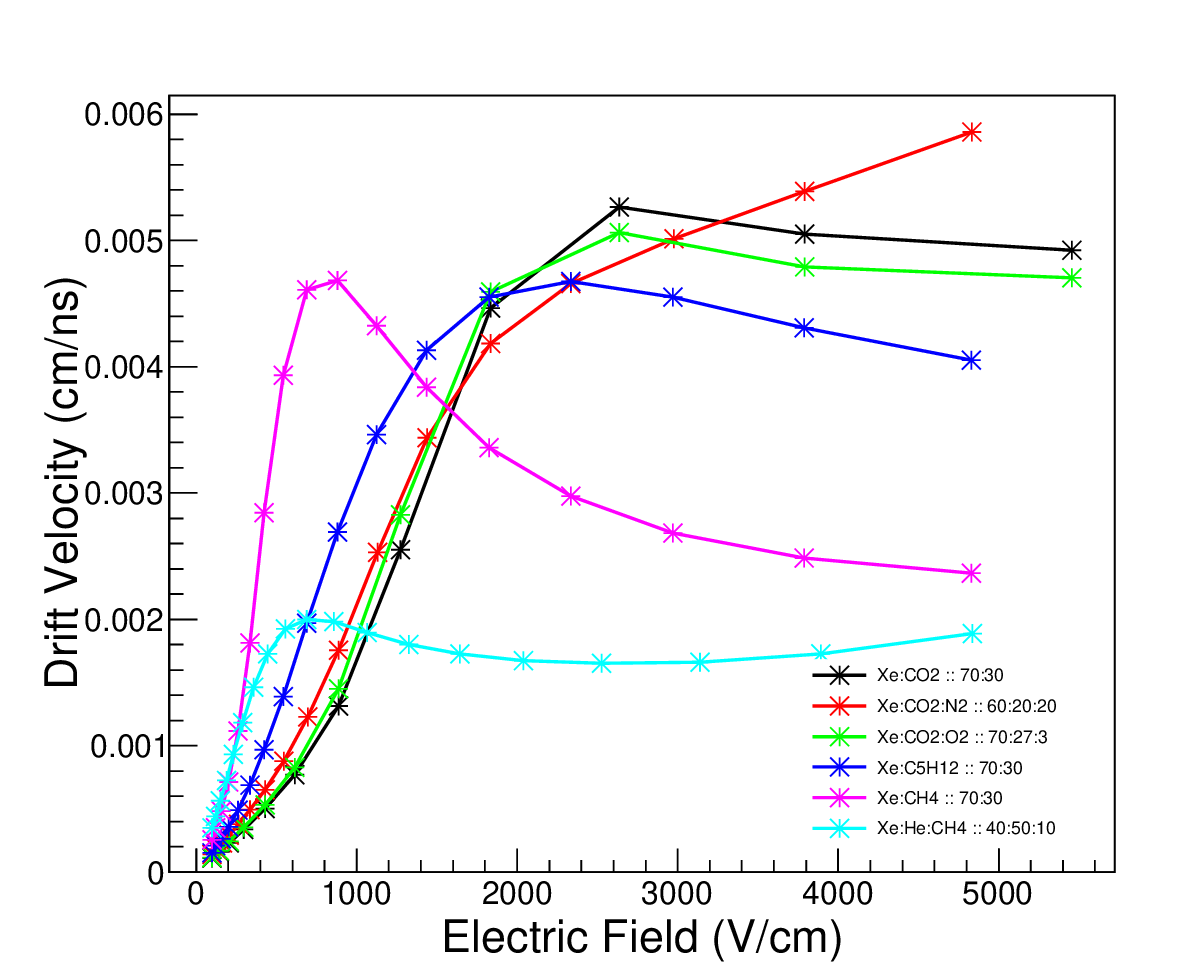}

\caption{The drift velocities of electrons in the ``selected'' gas mixtures as discussed in table~\ref{table21}. \label{fig:SevenOne}}
\end{figure}

\begin{figure}[htbp]
\centering
\includegraphics[width=0.8\textwidth]{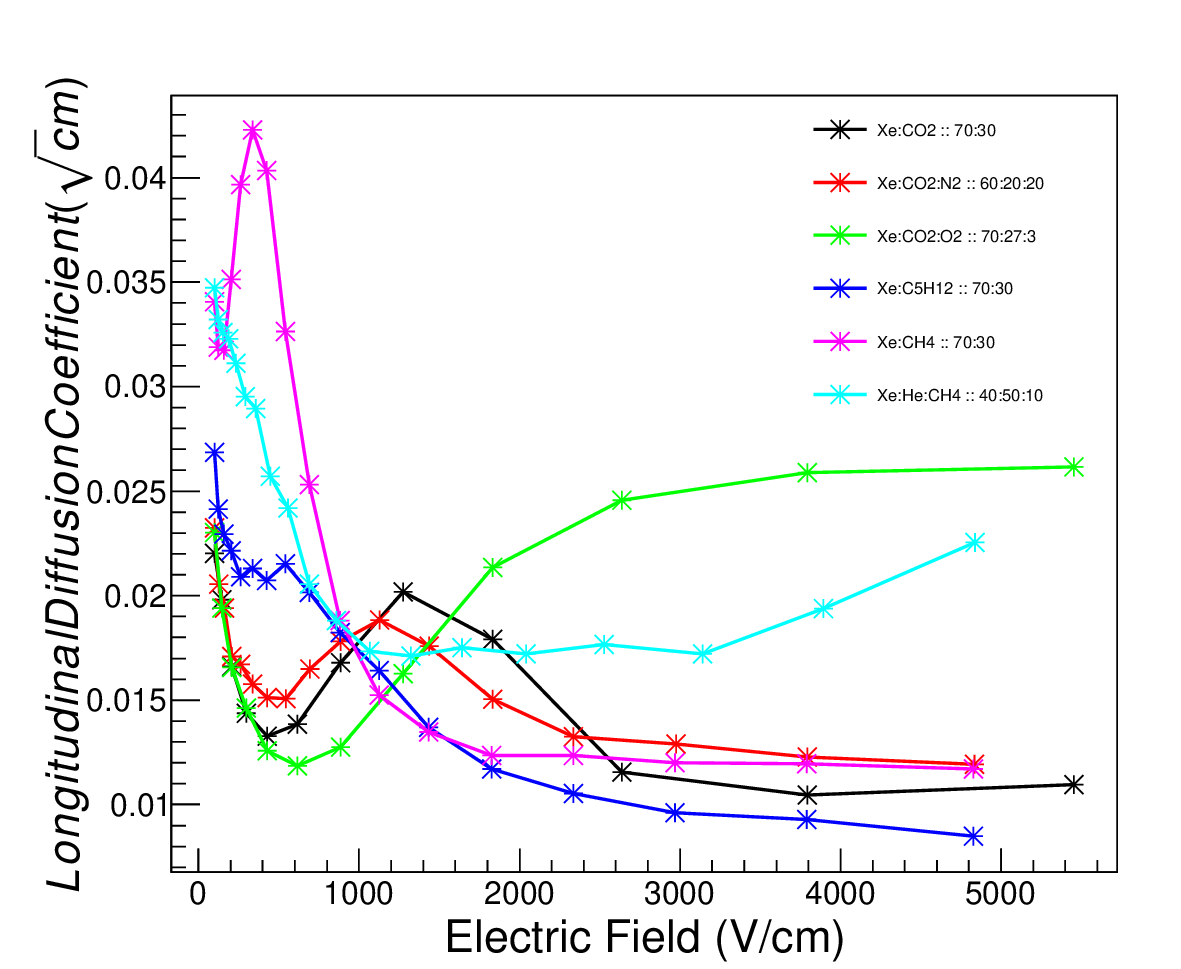}
\includegraphics[width=0.8\textwidth]{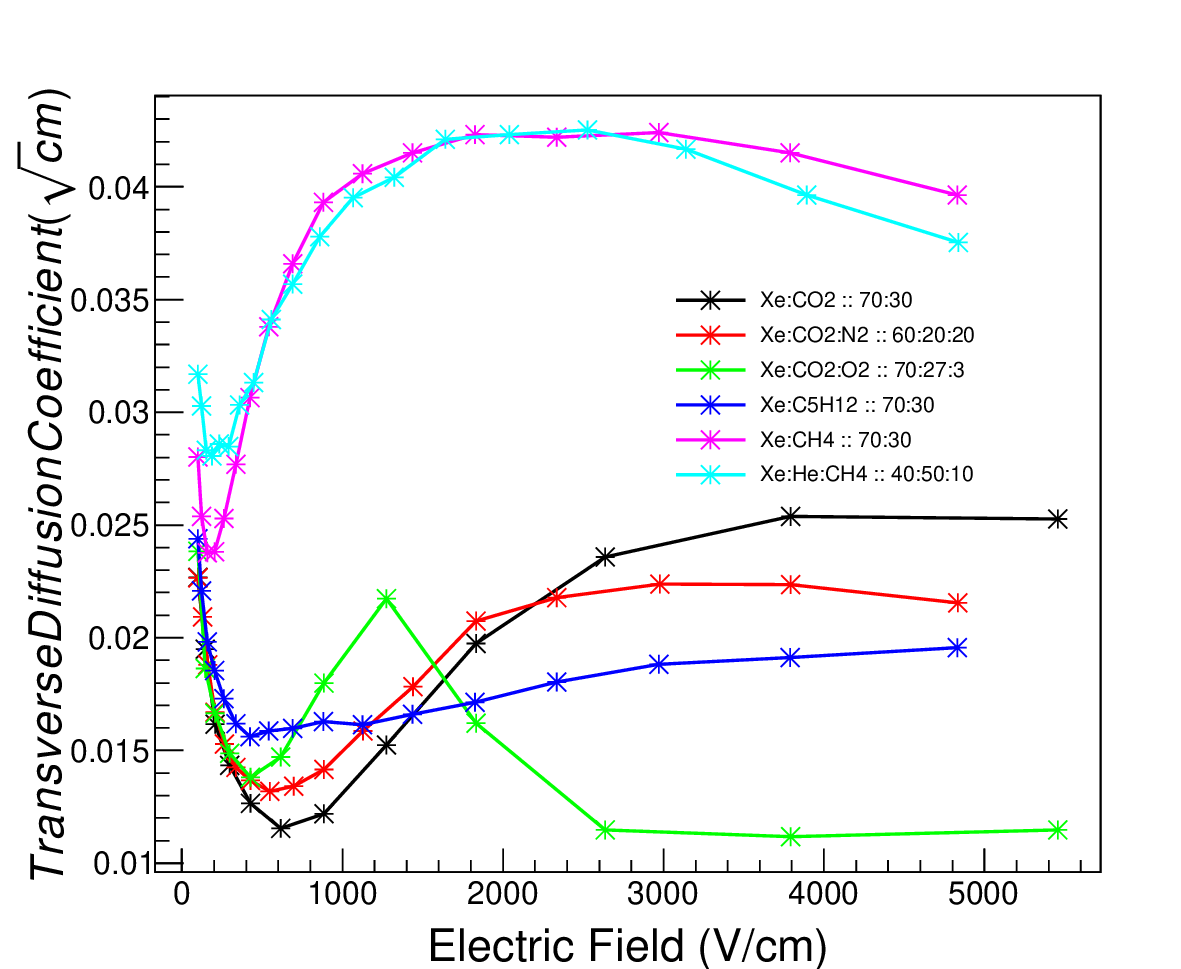}

\caption{The longitudinal and transverse diffusion coefficients in the ``selected'' gas mixtures as discussed in table~\ref{table21}. \label{fig:SevenTwo}}
\end{figure}

The ``selected'' gas mixtures as discussed in sub-section~\ref{PrimaryI} were analyzed to obtain the transportation of electrons in the gas mixtures when an electric field was applied. The Magboltz software \cite{gar} have been used for the study. Magboltz solves the Boltzmann transport equations in the gas mixtures under the application of electric and magnetic fields. With the application of electric field, the electrons drift along the field with the mean drift velocity. The drift velocities have been obtained for the ``selected'' gas mixtures as mentioned in the table~\ref{table21}. Figure~\ref{fig:SevenOne} shows the drift velocities of electrons and it increases with the increase in the electric field in a gas mixtures. It has been observed that $Xe:He:CH_{4}$ :: 40:50:10 gas mixture shows relatively low drift velocity. The electrons are drifted in the presence of the electric field. There are deviations of electrons from their average drift velocity due to scattering on the atoms of gas which leads to the fluctuations in the velocities that are measured as longitudinal and transverse diffusions. These longitudinal and transverse diffusion coefficients have been obtained with the ``selected'' gas mixtures as shown in figure~\ref{fig:SevenTwo}. It has been observed that $Xe:CH_{4}$ and $Xe:He:CH_{4}$ show relatively high transverse diffusion coefficients. The longitudinal diffusion coefficients have been observed to be higher for $Xe :CO_{2}:O_{2}$, whereas it is lowest for $Xe:C_{5}H_{12}$ at higher fields. In the next sub-section particle identification has been shown.

\subsection{Particle Identification with Straw Tube Detectors}

\begin{figure}[htbp]
  \centering
 
\includegraphics[width=1.\textwidth]{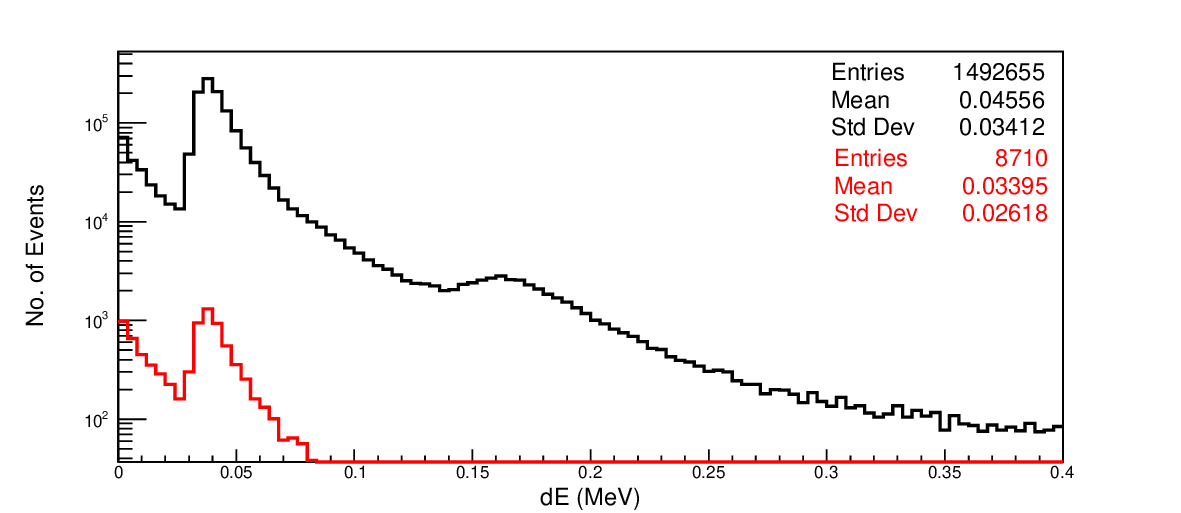}

\caption{The figure shows the energy loss distribution of pions and its decayed products when pions of 1 GeV energy were shot in the gas $XeCO_{2}$ :: 70:30. Note that, the color scheme for different decay products are: black - $\pi^{-}$, red - $\mu^{-}$.\label{fig:Nine}}
\end{figure}

The straw tube detector was also studied for the particle identification using different particles. This method of identification of different particles is independent of TR production and has been computed using ionization. Hence the $XeCO_{2}$ :: 70:30 gas mixture was selected for particle identification that was observed to show highest primary ionization as per the discussion in the sub-section~\ref{gasProp}. 10,000 pions were shot in the $XeCO_{2}$ :: 70:30 gas mixture. The physics lists chosen for the analysis were, G4HadronElasticPhysicsHP and G4HadronPhysicsFTFP\_BERT\_HP. The pions ($\pi^{-}$) that decays into muons ($\mu^{-}$) have been identified \cite{straw13}--\cite{straw23} in the gas mixture and has been shown in figure~\ref{fig:Nine}. 

\begin{figure}[htbp]
  \centering
\includegraphics[width=1.\textwidth]{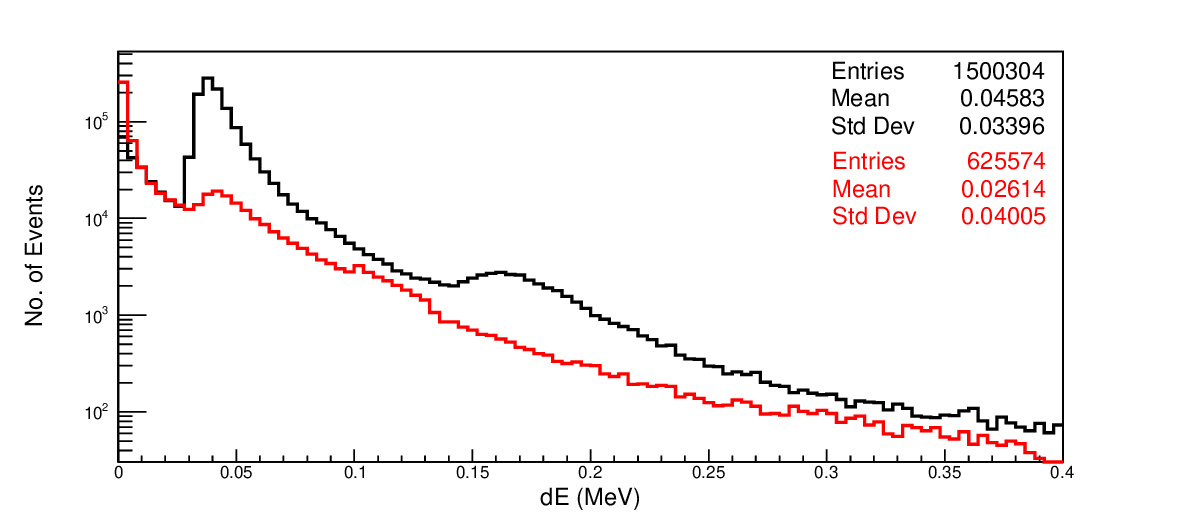}
\caption{The figure shows the energy loss distribution of muons (black) and its decayed products, electrons (red), when muons of 1 GeV energy were shot in the gas $XeCO_{2}$ :: 70:30. \label{fig:Ten}}
\end{figure}

Another particle that was chosen for particle identification was muons. 10,000 muons have been shot in the gas mixture $XeCO_{2}$ :: 70:30 for their identification. The physics list for the particle interactions was chosen to be G4EmStandardPhysics. The figure~\ref{fig:Ten} shows the muons that have been decayed into electrons in the gas mixture.

\begin{figure}[htbp]
  \centering
\includegraphics[width=1.\textwidth]{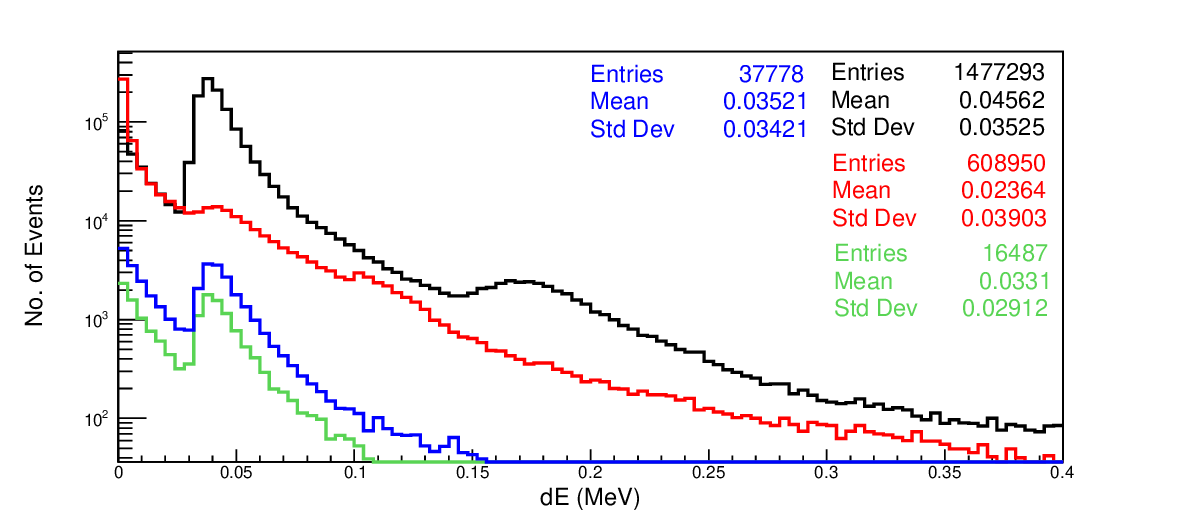}
\caption{The figure shows the energy loss distribution of kaons and its decayed products when kaons of 1 GeV energy were shot in the gas $XeCO_{2}$ :: 70:30. Note that, the color scheme for different products are: black - kaons, red - electrons, blue - $\mu^{-}$, green - $\pi^{-}$.\label{fig:Eleven}}
\end{figure}

The next particle chosen to be identified were kaons. 10,000 kaons were shot in the $XeCO_{2}$ :: 70:30 gas mixture. The physics lists were chosen to be same as that of pions. The kaons and its decay products have been identified in the detector as shown in figure~\ref{fig:Eleven}. Kaon decays into its several products as shown in the eq. \eqref{eqtwo}. These particles deposits their energies in the straw tube detector that has been shown in figure~\ref{fig:Eleven}. The kaon decays into $\mu^{-}$ and $\pi^{-}$ mainly, $\mu^{-}$ further decays into $e^{-}$ that has been shown in the figure. 

\begin{eqnarray}
\label{eqtwo}
%\begin{aligned}
K^{-} \rightarrow \mu^{-} + \bar\nu_{\mu}\\ \nonumber
K^{-} \rightarrow \pi^{-} + \pi^{0} \\ \nonumber
K^{-} \rightarrow \pi^{0} + \mu^{-} + \bar\nu_{\mu}\\ \nonumber
\mu^{-} \rightarrow e^{-} + \bar\nu_{e} \nonumber
%\end{aligned}
\end{eqnarray}

A summary of physics lists used in all the analysis has been given in table~\ref{table5}. In the next section the conclusions have been made.

\begin{table}[htbp]
%\centering
\caption{The physics lists used in all the analysis.\label{table5}}
\small
\begin{tabular}{|c|c|}
\hline
Physics Lists & Utilization \\
\hline
GammaNuclearPhysics and G4DecayPhysics & $^{60}$CO decay for gamma simulation \\
G4EmStandardPhysics & Electrons and Muons simulation \\
G4HadronElasticPhysicsHP and G4HadronPhysicsFTFP\_BERT\_HP & Pions, kaons simulation \\
\hline
\end{tabular}
\end{table}

\section{Summary and Conclusions}
\label{sec:summ}

The straw tube detectors are the type of detectors commonly used in high-energy physics experiments to track charged particles. They consist of thin-walled, gas-filled tubes (often made of a lightweight material) with a wire running along the central axis. When a charged particle passes through the tube, it ionizes the gas inside, creating free electrons and ions. The central wire, held at a high voltage, attracts the electrons, producing a measurable current. The straw tube detectors have wide range of applications due to their high precision, low mass and modularity.

In the present paper various studies have been done using different gas mixtures. The simulated STT detector is suitable for cancer therapy and various particle physics studies. The goals of the presented studies were primary ionization and particle identification. Additionally, the cobalt decay was studied. The results show that the energy spectrum during the decay of $^{60}$CO into $^{60}$Ni showed the two gamma peaks of 1.1 and 1.3 MeV energy respectively. The study with $^{60}$CO is done to explore the STT's potential towards the Cobalt therapy for cancer/tumor treatment. The different xenon-based gas mixtures i.e., $XeCO_{2}O_{2}$, $XeCO_{2}N_{2}$, $XeHeCH_{4}$, $XeCO_{2}$, $XeCH_{4}$, $XeC_{5}H_{12}$ have been utilized for the study of primary ionization and XTR photon production. The xenon-based gas mixtures are essential for the production of transition radiation and primary ionization. The straw tube detectors are the good detectors for the XTR study and posses good tracking capabilities. The various studies have been done after a lot of optimization of the gas mixtures. The XTR yield \cite{straw24}--\cite{straw25} was found to be highest in the $XeHeCH_{4}$::30:55:15 gas mixture. The primary ionization in the xenon-based gas mixtures have been done as it is essential property in the gaseous detectors. The number of primaries were found to be maximum in $XeCO_{2}$::70:30 gas mixture. The gas properties like the drift velocities and longitudinal and transverse diffusion coefficients have been obtained for studying the transportation of electrons when the electric field was applied. The particle identification has been done with the particles such as muons, pions and kaons in the straw tube detector. The performed studies can be useful for the future particle physics experiments to check the performance of the straw tube detectors.

\paragraph{Acknowledgements}: I would like to thank the department of physics, Chandigarh University for help and support.


\begin{thebibliography}{99}

\bibitem{compass}
V. N. Bychkov et. al., \emph{The large size straw drift chambers of the COMPASS experiment}, \emph{NIM A} {\bf 556} (1) (2006) 66-79. \href{https://doi.org/10.1016/j.nima.2005.10.026}{https://doi.org/10.1016/j.nima.2005.10.026}

\bibitem{pandas1} P. Gianotti et al., \emph{The straw tube trackers of the PANDA experiment}, \emph{$3^{rd}$ International Conference on Advancements in Nuclear Instrumentation, Measurement Methods and their Applications (ANIMMA), Marseille, France} (2013) 1--7. \href{https://doi.org/10.1109/ANIMMA.2013.6728039}{https://doi.org/10.1109/ANIMMA.2013.6728039}

\bibitem{pandas2} J. Smyrski et. al., \emph{Design of the forward straw tube tracker for the PANDA experiment}, \emph{JINST} {\bf 12} (2017) C06032. \href{https://doi.org/10.1088/1748-0221/12/06/C06032}{https://doi.org/10.1088/1748-0221/12/06/C06032}

\bibitem{pandas3} P. Wintz, \emph{For the PANDA Tracking Group}, \emph{The central straw tube tracker in the PANDA experiment}, \emph{Hyperfine Interactions} {\bf 229} (2014) 147--152. \href{https://doi.org/10.1007/s10751-014-1035-6}{https://doi.org/10.1007/s10751-014-1035-6}

\bibitem{atlas0} T. Akesson et.al., \emph{Straw tube drift-time properties and electronics parameters for the ATLAS TRT detector}, \emph{NIM A} {\bf 449} (3) (2000) 446-460. \href{https://doi.org/10.1016/S0168-9002(99)01470-9}{https://doi.org/10.1016/S0168-9002(99)01470-9}
  
\bibitem{atlas1} Ahmet Bingul, \emph{The ATLAS TRT and its Performance at LHC}, \emph{J. Phys.: Conf. Ser.} {\bf 347} (2012) 012025. \href{https://doi.org/10.1088/1742-6596/347/1/012025}{https://doi.org/10.1088/1742-6596/347/1/012025}
  
\bibitem{atlas2} Abat E. et al., \emph{The ATLAS TRT collaboration}, \emph{The ATLAS Transition Radiation Tracker (TRT) proportional drift-tube: Design and Performance, JINST} {\bf 3} (2008) P02013. \href{https://doi.org/10.1088/1748-0221/3/02/P02013}{https://doi.org/10.1088/1748-0221/3/02/P02013}
  
\bibitem{atlas3} Abat E. et al., \emph{The ATLAS TRT collaboration}, \emph{The ATLAS TRT Barrel Detector, JINST} {\bf 3} (2008) P02014. \href{https://doi.org/10.1088/1748-0221/3/02/P02014}{https://doi.org/10.1088/1748-0221/3/02/P02014}
  
\bibitem{atlas4} Abat E. et al., \emph{The ATLAS TRT collaboration}, \emph{The ATLAS TRT end-cap detectors, JINST} {\bf 3} (2008) P10003. \href{https://doi.org/10.1088/1748-0221/3/10/P10003}{https://doi.org/10.1088/1748-0221/3/10/P10003}
  
\bibitem{atlas5} Abat E. et al., \emph{The ATLAS TRT collaboration}, \emph{The ATLAS TRT electronics, JINST} {\bf 3} (2008) P06007. \href{https://doi.org/10.1088/1748-0221/3/06/P06007}{https://doi.org/10.1088/1748-0221/3/06/P06007} 

\bibitem{na62} A. Sergi, \emph{NA62 Spectrometer: A Low Mass Straw Tracker,
Physics Procedia} {\bf 37} (2012) 530--534. \href{https://doi.org/10.1016/j.phpro.2012.03.713}{https://doi.org/10.1016/j.phpro.2012.03.713}

\bibitem{gluex} N. S. Jarvis et. al., \emph{The Central Drift Chamber for GlueX. United States: N. p.} {\bf 962} (2020) 163727. \href{https://doi.org/10.1016/j.nima.2020.163727}{https://doi.org/10.1016/j.nima.2020.163727}

\bibitem{dune1} B. Abi et. al., \emph{Volume I. Introduction to DUNE, JINST} {\bf 15} (2020) T08008. \href{https://doi.org/10.1088/1748-0221/15/08/T08008}{https://doi.org/10.1088/1748-0221/15/08/T08008}

 \bibitem{dune2} A. A. Abud et. al., \emph{Deep Underground Neutrino Experiment (DUNE) Near Detector Conceptual Design Report, Instruments} {\bf 5}(4) (2021) 31. \href{https://doi.org/10.3390/instruments5040031}{https://doi.org/10.3390/instruments5040031}
 
\bibitem{straw1} T. Akesson et. al., \emph{Study of straw proportional tubes for a transition radiation detector/tracker at LHC, NIMA} {\bf 361} (3) (1995) 440--456. \href{https://doi.org/10.1016/0168-9002(95)00075-5}{https://doi.org/10.1016/0168-9002(95)00075-5}

\bibitem{rout} P. K. Rout, R. Kanishka et. al., \emph{Numerical estimation of discharge probability in GEM-based detectors, JINST} {\bf 16} (2021) P09001. \href{10.1088/1748-0221/16/09/P09001}{10.1088/1748-0221/16/09/P09001}
  
\bibitem{straw2} T. Akesson et. al., \emph{ATLAS TRT collaboration}, \emph{Electron identification with a prototype of the Transition Radiation Tracker for the ATLAS experiment, NIMA} {\bf 412} (2--3) (1998) 200--215. \href{https://doi.org/10.1016/S0168-9002(98)00457-4}{https://doi.org/10.1016/S0168-9002(98)00457-4}

\bibitem{geant4} S. Agostinelli et al., \emph{Geant4-a simulation toolkit, NIMA} {\bf 506}, (2003)250. \href{https://geant4.web.cern.ch/}{https://geant4.web.cern.ch/}

\bibitem{root} \href{https://root.cern/}{https://root.cern/}.

\bibitem{straw3} Adrian Vogel, \emph{For ATLAS Collaboration, ATLAS Transition Radiation Tracker (TRT): Straw tube gaseous detectors at high rates, NIMA} {\bf 732} (2013) 277--280. \href{https://doi.org/10.1016/j.nima.2013.07.020}{https://doi.org/10.1016/j.nima.2013.07.020} %70% Xe + 27% CO2 + 3% O2

\bibitem{straw4} Bartosz Mindur, \emph{For ATLAS Collaboration, ATLAS Transition Radiation Tracker (TRT): Straw tubes for tracking and particle identification at the Large Hadron Collider, NIMA} {\bf 845} (2017) 257--261. \href{https://doi.org/10.1016/j.nima.2016.04.026}{https://doi.org/10.1016/j.nima.2016.04.026} %70% Xe + 27% CO2 + 3% O2

\bibitem{straw5} G. D. Barr et. al., \emph{A large-area transition radiation detector, NIMA} {\bf 294}(3) (1990) 465--472 \href{https://doi.org/10.1016/0168-9002(90)90287-G}{https://doi.org/10.1016/0168-9002(90)90287-G} %Xe + 55% He + 15% CH4

\bibitem{straw6} K. Aamodt et. al., \emph{For The ALICE Collaboration, The ALICE experiment at the CERN LHC, JINST} {\bf 3} (2008) S08002. \href{https://doi.org/10.1088/1748-0221/3/08/S08002}{https://doi.org/10.1088/1748-0221/3/08/S08002}  %xeco2  
  
\bibitem{straw7} B. Libby et. al., \emph{Particle identification in TEC/TRD Prototypes for the PHENIX detector at RHIC, NIMA} {\bf 367}(1--3) (1995) 244--247. \href{https://doi.org/10.1016/0168-9002(95)00646-X}{https://doi.org/10.1016/0168-9002(95)00646-X}  %xech4
  
\bibitem{straw8} Boris Dolgoshein, \emph{Transition radiation detectors, NIMA} {\bf 326}(3) (1993) 434--469. \href{https://doi.org/10.1016/0168-9002(93)90846-A}{https://doi.org/10.1016/0168-9002(93)90846-A} %Xec5h12
  
\bibitem{straw12b} A. Andronic, J. P. Wessels, \emph{Transition radiation detectors, NIMA} {\bf 666} (2012) 130--147. \href{https://doi.org/10.1016/j.nima.2011.09.041}{https://doi.org/10.1016/j.nima.2011.09.041} %Main
  
\bibitem{cobalt} P. Rice-Evans and Z. Aung, \emph{On the decay of cobalt 60, Z. Physik} {\bf 240} (1970) 392--395. \href{https://doi.org/10.1007/BF01395575}{https://doi.org/10.1007/BF01395575}

\bibitem{straw9} M. L. Cherry et. al., \emph{Transition radiation from relativistic electrons in periodic radiators, Phys. Rev. D} {\bf 10}(11) (1974) 3594--3607. \href{https://doi.org/10.1103/PhysRevD.10.3594}{https://doi.org/10.1103/PhysRevD.10.3594} %TR

\bibitem{straw10} X. Artru et. al., \emph{Practical theory of the multilayered transition radiation detector, Phys. Rev. D} {\bf 12} (1975) 1289--1306. \href{https://doi.org/10.1103/PhysRevD.12.1289}{https://doi.org/10.1103/PhysRevD.12.1289} %TR
  
\bibitem{straw11} L. Durand, \emph{Transition radiation from ultrarelativistic particles, Phys. Rev. D} {\bf 11}(1) (1975) 89--105. \href{https://doi.org/10.1103/PhysRevD.11.89}{https://doi.org/10.1103/PhysRevD.11.89}   %TR

\bibitem{straw12} E. Barbarito et. al., \emph{A large area transition radiation detector to measure the energy of muons in the Gran Sasso underground laboratory, NIMA} {\bf 365}(1) (1995) 214--223 \href{https://doi.org/10.1016/0168-9002(95)00475-0}{https://doi.org/10.1016/0168-9002(95)00475-0}  %TR

\bibitem{straw12a} J. Gu et. al., \emph{The photon yield efficiency study of transition radiators at E2 line of Beijing Test Beam Facility, JINST} {\bf 16} (2021) P08041 \href{https://doi.org/10.1088/1748-0221/16/08/P08041}{https://doi.org/10.1088/1748-0221/16/08/P08041}

\bibitem{primary1} R. Kanishka, S. Mukhopadhyay, N. Majumdar, S. Sarkar, \emph{A
  Simulation of Primary Ionization for Different Gas Mixtures, Springer Proc. Phys.} {\bf 282} (2023) 47--53. \href{https://doi.org/10.1007/978-3-031-19268-5\_6}{https://doi.org/10.1007/978-3-031-19268-5\_6}

\bibitem{primary2} R. Kanishka et. al., \emph{Primary Ionization Simulation for Different Gas Mixtures, J. Phys.: Conf. Ser.} {\bf 2349} (2022) 012019. \href{https://doi.org/10.1088/1742-6596/2349/1/012019}{https://doi.org/10.1088/1742-6596/2349/1/012019}
  
\bibitem{gar} \href{https://magboltz.web.cern.ch/magboltz/}{https://magboltz.web.cern.ch/magboltz/}
  
\bibitem{straw13} C. W. Fabjan, W. Struczinkski, \emph{Coherent emission of transition radiation in periodic radiators, Phys. Lett. B} {\bf 57}(5) (1975) 483--486. \href{https://doi.org/10.1016/0370-2693(75)90274-9}{https://doi.org/10.1016/0370-2693(75)90274-9}
  
\bibitem{straw14} C. Camps et al., \emph{Transition Radiation from Electrons at Low gamma Values, Nucl. Instr. and Meth.} {\bf 131}(3) (1975) 411--416. \href{https://doi.org/10.1016/0029-554X(75)90426-7}{https://doi.org/10.1016/0029-554X(75)90426-7}
  
%\bibitem{straw15} T. A. Prince et al., Nucl. Instr. and Meth. 123, 231 (1975).
  
\bibitem{straw16} J. Cobb et. al., \emph{Transition Radiators for electron Identification at the CERN ISR, Nucl. Instr. and Meth.} {\bf 140}(3) (1977) 413--427. \href{https://doi.org/10.1016/0029-554X(77)90355-X}{https://doi.org/10.1016/0029-554X(77)90355-X}
  
\bibitem{straw17} C. W. Fabjan et. al., \emph{Transition Radiation Spectra from Randomly Spaced Interfaces, Nucl. Instr. and Meth.} {\bf 146}(2) (1977) 343--346. \href{https://doi.org/10.1016/0029-554X(77)90718-2}{https://doi.org/10.1016/0029-554X(77)90718-2}
  
\bibitem{straw18} M. L. Cherry et. al., \emph{Measurements of the spectrum and energy dependence of x-ray transition radiation, Phys. Rev. D} {\bf 17} (1978) 2245. \href{https://doi.org/10.1103/PhysRevD.17.2245}{https://doi.org/10.1103/PhysRevD.17.2245}
  
\bibitem{straw19} C. W. Fabjan et. al., \emph{Practical Prototype of a Cluster Counting Transition Radiation Detector, Nucl. Instr. and Meth.} {\bf 185}(1--3) (1981) 119--124. \href{https://doi.org/10.1016/0029-554X(81)91202-7}{https://doi.org/10.1016/0029-554X(81)91202-7}
  
\bibitem{straw20} A. Bungener et. al., \emph{ELECTRON IDENTIFICATION BEYOND 1-GEV BY MEANS OF TRANSITION RADIATION, Nucl. Instr. and Meth.} {\bf 214}(2--3) (1983) 261--268. \href{https://doi.org/10.1016/0167-5087(83)90592-6}{https://doi.org/10.1016/0167-5087(83)90592-6}
  
\bibitem{straw21} R. D. Appuhn et. al., \emph{Transition Radiation Detectors for Electron Identification Beyond 1-GeV/c, NIMA} {\bf 263}(2--3) (1988) 309--318 \href{https://doi.org/10.1016/0168-9002(88)90965-5}{https://doi.org/10.1016/0168-9002(88)90965-5}

\bibitem{straw22} V. Commichau et. al., \emph{A transition radiation detector for pion identification in the 100 GeV/c momentum region, Nuclear Instruments and Methods} {\bf 176}(1--2) (1980) 325--331. \href{https://doi.org/10.1016/0029-554X(80)90724-7}{https://doi.org/10.1016/0029-554X(80)90724-7}

\bibitem{straw23} M. Deutschmann et. al., \emph{Particle identification using the angular distribution of transition radiation, Nuclear Instruments and Methods} {\bf 180}(2--3) (1981) 409--412 \href{https://doi.org/10.1016/0029-554X(81)90080-X}{https://doi.org/10.1016/0029-554X(81)90080-X}
%%%%


\bibitem{straw24} M. L. Cherry, \emph{Measuring the Lorentz Factors of Energetic Particles with Transition Radiation, Nucl. Instrum. Meth. in Phys. Res. A} {\bf  706} (2013) 39--42. \href{https://doi.org/10.1016/j.nima.2012.05.008}{https://doi.org/10.1016/j.nima.2012.05.008}

\bibitem{straw25} M. Albrow et. al., \emph{Transition radiation detectors for hadron separation in the forward direction of LHC experiments, NIMA} {\bf 1055} (2023) 168535. \href{https://doi.org/10.1016/j.nima.2023.168535}{https://doi.org/10.1016/j.nima.2023.168535}

  
  
  
\end{thebibliography}
\end{document}